\documentclass[journal=jacsat,manuscript=article]{achemso}

\usepackage[version=3]{mhchem} 
\usepackage{amsmath} 
\usepackage{bm}  
\usepackage{amsmath}
\usepackage{amssymb} 
\usepackage{multirow} 
\usepackage{mathtools} 
\usepackage{physics}   
\usepackage{gensymb} 
\usepackage{graphicx} 
\usepackage{adjustbox} 
\usepackage{natbib} 
\usepackage{isotope} 
\usepackage{nameref} 
\usepackage{booktabs} 
\usepackage{lscape} 
\usepackage{subcaption} 
\usepackage{xcolor} 
\usepackage{comment} 

\newcommand{\water}{\ce{H2^{\!17}O}}

\newcommand{\ham}{\hat{\cal H}} 

\newcommand\Tstrut{\rule{0pt}{2.6ex}}         
\newcommand\Bstrut{\rule[-0.9ex]{0pt}{0pt}}   

\author{Shiva Agarwal}
\affiliation[Western Michigan University]
{Department of Physics, Western Michigan University, Kalamazoo, MI, 49008, USA}
\author{Sungsool Wi}
\affiliation[National High Magnetic Field Laboratory]
 {National High Magnetic Field Laboratory, 
Tallahassee, FL, 32310, USA}
\author{Jason Kitchen}
\affiliation[National High Magnetic Field Laboratory]
 {National High Magnetic Field Laboratory, 
Tallahassee, FL, 32310, USA}
\author{Zhongrui Li}
\affiliation[University of Michigan]
{Electron Microbeam Analysis Laboratory, 
University of Michigan, Ann Arbor, MI, 48109, USA}
\author{Christopher J. Taylor}
\affiliation[Western Michigan University]
{Department of Chemistry, Western Michigan University, Kalamazoo, MI, 49008, USA}
\author{Michael A. Famiano}
\affiliation[Western Michigan University]
{Department of Physics, Western Michigan University, Kalamazoo, MI, 49008, USA}
\author{John B. Miller}
\email{john.b.miller@wmich.edu}
\phone{+1 (269) 387-2871}
\affiliation[Western Michigan University]
{Department of Chemistry, Western Michigan University, Kalamazoo, MI, 49008, USA}

\title[Magnetic shielding in enantiomers]
  {Single-Crystal NMR for \ce{\mathbf{^{17}}O} in Alanine Enantiomers}

\abbreviations{NMR}
\keywords{amino acids, enantiomers, chirality, homochirality, Single-crystal NMR spectroscopy, shielding tensors, \isotope[17]{O}, DFT, x-ray diffraction}

\begin{document}

\begin{tocentry}
\begin{center}
    \includegraphics[width=\linewidth]{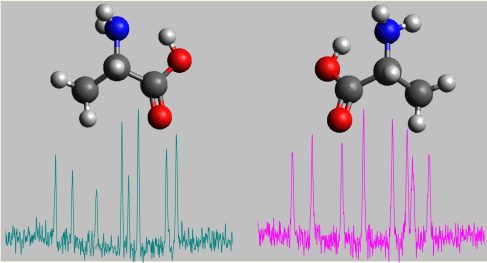}\\
    \vspace{0.5ex}
    \textbf{For Table of Contents Only}
\end{center}

\end{tocentry}

\begin{abstract}
\noindent 
Single-crystal solid-state nuclear magnetic resonance (ssNMR) spectroscopy, which enables detailed analysis of the electronic structures of crystalline molecules, offers a unique opportunity to investigate molecular chirality---an essential feature with broad implications for understanding the origin and function of life. In this study, we employ single-crystal ssNMR spectroscopy, in combination with X-ray diffraction and density functional theory (DFT) calculations, to examine the electronic structure of \isotope[17]{O} nuclei in crystalline forms of alanine enantiomers. Eight magnetically nonequivalent \isotope[17]{O} resonances within the unit cell were observed and successfully assigned, and their corresponding NMR tensor parameters were determined. The experimental findings were compared with previous NMR studies as well as with DFT calculations performed in this work. The DFT results not only supported the assignment of crystallographically distinct \isotope[17]{O} sites but also revealed previously unobserved antisymmetric components of the chemical shift tensors. This study presents the first comprehensive characterization of \isotope[17]{O} NMR tensors in alanine enantiomers and underscores the power of integrating single-crystal ssNMR with X-ray diffraction and DFT calculations to advance our understanding of molecular chirality in amino acids.
\end{abstract}

\section{Introduction}
Biological homochirality is a fundamental feature of  terrestrial life~\cite{chen2020origin, blackmond_origin_2019}. Analysis of a number of carbonaceous meteorites have revealed an excess of \textit{L-}~amino acids compared to the \textit{D-}~enantiomers~\cite{glavin_extraterrestrial_2021, burton_understanding_2012, elsila_meteoritic_2016}. These findings suggest that abiotic mechanism(s) operating in the  stellar environments may be responsible for generating this enantiomeric excess (\textit{ee})~\cite{burton_insights_2018}. Several abiotic models~\cite{ulbricht_attempts_1962, garcia_astrophysical_2019,  thiemann_speculations_1984, quack_how_2002, famiano_chiral_2021} have been proposed to explain the observed \textit{ee} of amino acids. Among them, the magnetochiral model calculates \textit{ee} of as high as 0.02\% for alanine, positive \textit{ee} for many $\alpha$-amino acids, and up to 0.01\% for cationic isovaline and zwitterionic alanine~\cite{famiano_selection_2018, famiano_amino_2018, famiano_chiral_2021}.

The orientation dependence arising from optical isomerism influences the relative nuclear interaction rates of chiral \isotope[14]{N} nuclei in amino acids with relativistic leptons, such as electron antineutrinos ($\bar\nu_e$), potentially leading to their conversion into \isotope[14]{C}. This mechanism may contribute to the preferential destruction of \textit{D-}~amino acids over their \textit{L-}~counterparts. In single-crystal ssNMR experiments, the orientation of chiral molecules in high magnetic fields alters the electronic environment such that the antisymmetric components of the magnetic shielding tensor are altered to exhibit reflection symmetry. 

The initial motivation of this work was to experimentally test the predictions of the magnetochiral model. The primary aim was to examine the antisymmetric chemical shielding (ACS) tensor components of the \isotope[14]{N} nucleus ($I = 1$, $\gamma_{\isotope[14]{N}} =  1.934 \times 10^7\ \text{rad}\ \text{T}^{-1}\ \text{s}^{-1}$) in amino acid samples. Although ACS contributions have only been indirectly inferred through relaxation studies in solution-state NMR~\cite{paquin_determination_2010, anet1990nmr}, for quadrupolar nuclei ($I>1/2$), the quadrupolar and ACS components are coupled to each other and can potentially be measured using solid-state NMR spectroscopy~\cite{wi_quadrupolar-shielding_2002}. Unfortunately, our efforts to observe \isotope[14]{N} nucleus in chiral systems have thus far been unsuccessful. This is primarily due to  its low gyromagnetic ratio, which results in poor sensitivity in NMR spectroscopy. Additionally, the nucleus has integer spin ($I = 1$), leading to two non-symmetrical transitions ($\ket{1}\leftrightarrow\ket{0} \text{and\ } \ket{0}\leftrightarrow\ket{-1}$), and a relatively large quadrupolar coupling constant ($C_Q$). These factors contribute to line broadening and signal loss,  making \isotope[14]{N} particularly challenging to detect under a moderate field strength, 14.1 T~\cite{veinberg201614n}.

Instead, we turned to the oxygen atoms in the carboxylate moiety of the amino acids.  Of the three stable oxygen isotopes: \isotope[16]{O},  \isotope[17]{O}, and  \isotope[18]{O}, only \isotope[17]{O} ($I = 5/2$) is NMR active. Being a quadrupolar nucleus as well, \isotope[17]{O} can serve as a proxy for the \isotope[14]{N} as a probe of the chiral environments of most amino acids. 
Due to its low natural abundance ($\sim~0.04\%$), isotopically enriched samples are required for rapid and facile \isotope[17]{O} NMR spectroscopy experiments~\cite{spackova_fast_2023}.

In this work, we have investigated the quadrupolar tensor and the chemical shielding anisotropy (CSA) tensor for \isotope[17]{O} nuclei in both enantiomers of isotopically-enriched alanine. The relative orientations of these tensors were determined through single-crystal NMR spectroscopy. The experimental findings were compared with previously reported NMR studies on alanine enantiomers~\cite{yamauchi_17o_1999, pike_solid-state_2004, wu_two-dimensional_2001}, as well as with periodic density functional theory (DFT) calculations using the published crystal structure~\cite{wilson_neutron_2005}. The DFT models were performed using the PBEsol exchange-correlation functionals~\cite{perdew_restoring_2008}. The NMR tensors  were calculated using the gauge-including projector augmented wave (GIPAW) method~\cite{GIPAW}.
We also present similar calculated data for the \isotope[14]{N} nuclei.

\section{Theory}
The NMR rotating frame Hamiltonian of an isolated quadrupolar nucleus, considered in the Zeeman interaction frame includes contributions from chemical shielding and quadrupolar interactions and is expressed as~\cite{haeberlen_1976}:
\begin{equation}
    \ham_{Rot}^{Total} = \ham_Q^{(1)} + \ham_Q^{(2)} + \ham_{CSA}^{(1)}
    \label{eq: nmr rotating hamiltonian}
\end{equation}

with

\begin{align}
    \ham_Q^{(1)} &= \chi_QR_{2,0}^Q\qty{3I_z^2 - I\qty(I+1)} & & \qty( \because \chi_Q = \frac{eQ}{2I(2I - 1)\hbar}), \label{eq:HQ1}\\
    \begin{split}
  \ham_Q^{(2)} &= \frac{1}{2\omega_0} \chi_Q^2 \bigg[ R_{2,-1}^Q R_{2,1}^Q I_z \qty{4I(I+1) - 8I_z^2 - 1} \\
  &\qquad+ R_{2,-2}^Q R_{2,2}^Q I_z \qty{2I(I+1) - 2I_z^2 - 1} \bigg]\label{eq:HQ2}
\end{split}
\end{align}
and
\begin{align}
\ham_{CSA}^{(1)} = \qty{\delta_{\text{iso}} + R_{2,0}^{CSA}}\gamma B_o I_z \label{eq:HCSA1} 
\end{align}

In the above equations, $\ham_Q^{(1)}, \ham_Q^{(2)} , \ham_{CSA}^{(1)}$ represent the first order quadrupolar interaction, second-order quadrupolar interaction, and first order chemical shift interaction, respectively.  The term $\delta_{\text{iso}}$ denotes the isotropic chemical shift, and the components $R_{2,\lambda}^\xi$ (with $\xi$=CSA or Q, and $\lambda$=2, 1, 0, -1, or -2) represent the spatial part of tensors defined in the laboratory (rotating) frame. Here, $I$ is the nuclear spin quantum number, $I_z$ is the z-component of the angular momentum operator, and $eQ$ is the nuclear quadrupole moment, $\gamma$ is the gyromagnetic ratio, and $\omega_0$ is the nuclear Larmor frequency. The spherical tensors $R_{2,\lambda}^\xi$, defined in the laboratory frame, can be related to the corresponding $G_{2,\lambda}^\xi$ tensors defined in the goniometer-tenon frame through a single-step coordinate transformation using the polar angle $\theta$ and an azimuthal angle $\phi$ according to~\cite{vosegaard_single-crystal_2021, wi_quadrupolar-shielding_2002}
\begin{align}
    {\text{Goniometer frame}} 
    \xrightarrow{(\phi, \theta, 0^\circ)} 
    \text{Laboratory frame} & & (B_o\ \text{along the $z$-axis} )
    \label{eq:frames}
\end{align}
After explicitly performing this transformation and re-expressing the spherical tensors  $G_{2,\lambda}^\xi$ into Cartesian tensors $G_{\text{mn}}^\xi$ (where m and n are $x$, $y$, or $z$) for easier interpretation, the expression for $R_{2,\lambda}^\xi\ \text{and}\ R_{2,\lambda}^\xi R_{2,\lambda'}^\xi$ can be written as functions of $G_{\text{mn}}^\xi$, $\theta$, and $\phi$, as described previously~\cite{agarwal2025boron}.

As shown in Figure~\ref{fig:probe} below, the goniometer tenon, to which the sample crystal is glued for mounting in the single-crystal NMR probe, is designed to allow rotations about the -$x$, $y$, and -$z$ axes. The relevant expressions used to interpret the -$x^T$ rotation ($\theta = -\Theta; \phi = \pi/2$), $y^T$ rotation ($\theta = \Theta; \phi = 0$), and -$z^T$ rotation ($\theta = \pi/2; \phi= -\Theta$) patterns for the transitions of \isotope[17]{O} ($I = 5/2$) can be derived from the Eqs.~\ref{eq:HQ1} -~\ref{eq:HCSA1} above by considering transitions among different energy levels. Here, $\Theta$ represents the rotation angle applied experimentally by rotating the crystal about an axis that is oriented $90^\circ$ relative to the external magnetic field (see Figure~\ref{fig:probe}). For the central transition $\ket{1/2} \leftrightarrow \ket{-1/2}$, all three rotation patterns can be rearranged into the following expressions for explicit curve fitting, incorporating the first-order CSA and second-order quadrupolar contributions:

\begin{align}
    \nu^{CSA}_{\ket{1/2}\leftrightarrow\ket{-1/2}}  &= A^{CSA} + B^{CSA}\cos  2\Theta + C^{CSA} \sin  2\Theta
    \label{eq: CS fit equation}\\
    \nu^{Q2}_{\ket{1/2}\leftrightarrow\ket{-1/2}}  &= A^{Q} + B^{Q}\cos 2\Theta + C^{Q} \sin  2\Theta + D^{Q}\cos 4\Theta + E^{Q} \sin 4\Theta
    \label{eq: Q fit equation}
\end{align}
where $\alpha \in \{\text{-}x^T, y^T, \text{-}z^T\}$ and the coefficients $\Gamma_m^\xi$ (with $\Gamma = A, B, C, \dotso, E $ ) are defined in terms of the $G_{\text{mn}}^\xi$ tensors,~\cite{agarwal2025boron} which are  (3×3 matrices). Note that the dominant first-order quadrupolar Hamiltonian vanishes for symmetric transitions such as $\ket{1/2} \leftrightarrow \ket{-1/2}$, due to its quadratic dependence on $I_z^2$. 

The CSA and quadrupolar tensor parameters $G_{\text{mn}}^\xi$ in the goniometer–tenon frame are determined by performing least-squares curve fitting of the experimentally acquired -$x^T$, $y^T$, and -$z^T$ rotation patterns using the expressions above~\cite{vosegaard_quadrupole_1996}. The tensor parameters in the respective principal axis frames (PAFs) of the CSA and quadrupolar tensors are then obtained via matrix diagonalization. The corresponding eigenvector matrices yield the relative orientations of each tensor with respect to the goniometer–tenon frame. These principal components represent unique molecular properties and provide a generalized framework for describing the corresponding NMR tensor interactions. In the principal axis frame (PAF), the CSA and quadrupolar tensors can be represented as follows:

\begin{align}
    \delta_{\text{iso}} &= \frac{1}{3}({\delta_{11} + \delta_{22} + \delta_{33}})
    \label{eq:  CS iso}\\
    \delta_{\text{CS}} &= \delta_{33} - \delta_{\text{iso}}
    \label{eq: delta CS}\\
    \eta_{\text{CS}} &= \frac{\delta_{22} - \delta_{11}}{\delta_{\text{CS}}}
    \label{eq: eta CS}
\end{align}
with the convention~\cite{haeberlen_1976} 
\begin{align}
    |\delta_{33} - \delta_{\text{iso}}| \geq |\delta_{11} - \delta_{\text{iso}}| \geq |\delta_{22} - \delta_{\text{iso}}|  \label{eq: CS order}
\end{align}
for CSA and
\begin{align}
       C_Q &= \frac{eQ\cdot V_{33}}{h}
    \label{eq: delta Quadrupolar}\\
    \eta_Q &= \frac{V_{22} - V_{11}}{V_{33}}
    \label{eq: eta Quadrupolar}  
\end{align}
with the convention
\begin{align}
    |V_{33}| \geq |V_{11}| \geq |V_{22}| \label{eq: CQ order}
\end{align}
for the quadrupolar tensors. Here, $\delta_{\text{mn}}^{CS}$ and $V_{\text{mn}}^Q$ denote CSA tensor and electric field gradient (EFG) tensor, respectively, each defined in its corresponding PAF. 

Tensors defined in the principal axis frame (PAF) and the goniometer–tenon frame are connected via a unitary transformation. In most cases, the PAF of the quadrupolar interaction is chosen as the common reference frame, enabling expression of the CSA tensor’s PAF—--and when applicable, the crystal frame from x-ray diffraction—--in the same coordinate system. When the crystal frame is adopted as the common reference, both the quadrupolar and CSA PAFs can be mathematically converted into the goniometer–tenon frame, which corresponds to the physical orientation of the mounted crystal, using the following tensor transformations each consisting of three consecutive passive rotations involving an Euler’s angle set ($\alpha_1, \beta_1, \gamma_1$):

\begin{equation}
    \text{PAF(CSA)} 
    \xrightarrow{(a, b, c)} 
    \text{PAF(Q)} 
    \xrightarrow{(\zeta, \lambda, \nu)} 
    \text{Crystal frame}
    \xrightarrow{(\alpha, \beta, \gamma)} 
    \underset{\text{(Rotating frame)}}{\text{Goniometer frame}} 
    \label{eq:frames}
\end{equation}
Thus the tensors in each frame, (e.g., A and B) are related by a unitary transformation as~\cite{millot2012active}:
\begin{equation}
    B = R(\alpha_1, \beta_1, \gamma_1)AR^{-1}(\alpha_1, \beta_1, \gamma_1) \label{eq: Euler orientation}
\end{equation}

\section{Experimental Details}

\subsection{\isotope[17]{O} Labeling}

The colorless \textit{D-} and \textit{L-} enantiomers of  alanine  were individually labeled by a saponification reaction of the corresponding alanine methyl ester hydrochloride (Ala-OMe$\cdot$HCl) with sodium ethoxide (\ce{NaOEt}) and isotopically-enriched water (\water), following a procedure previously described~\cite{spackova_fast_2023}. Both enantiomers of Ala-OMe$\cdot$HCl and \ce{NaOEt} were purchased from Sigma-Aldrich while \water\ with nominal $40\%$ enrichment was purchased from Cambridge Isotope Laboratories (CIL).  The relevant chemical reaction is given in Scheme~\ref{Scheme: Labeling}. 
\begin{scheme}[h]
    \centering
   \caption{\isotope[17]{O} labeling of H-Ala-OMe$\cdot$HCl}
    \includegraphics[width=\textwidth]{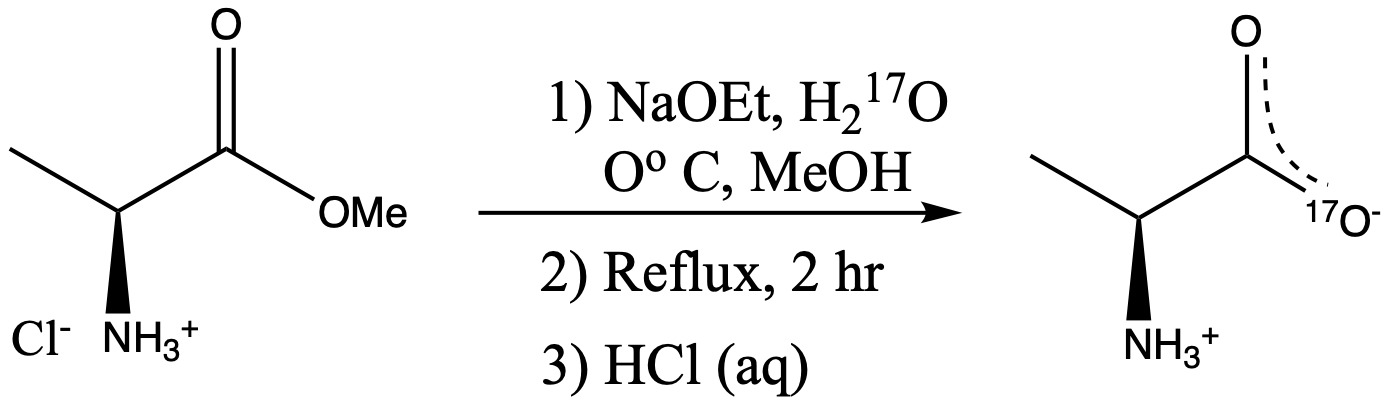}
\label{Scheme: Labeling}
\end{scheme}
Incorporation of the \isotope[17]{O} label in the two alanine enantiomers was measured to be similar to the predicted 20 atom-\% label by solution NMR, where the intensity of the \textit{ca.} 265 ppm carboxylate resonance was referenced to that of the natural abundance \isotope[17]{O} of the \ce{D2O} solvent. Mass spectrometric analysis confirmed the isotope incorporation for each labeled compound (see Supplementary Material). The \isotope[1]{H} and \isotope[13]{C} NMR spectra, all collected in \ce{D2O}, were consistent with previous literature~\cite{pouchert1993aldrich}.

Single crystals for both \textit{D-} and \textit{L-}alanine were prepared separately by slow evaporation at room temperature after dissolving $\sim1.5$ g of product in $\sim9$ ml cold deionized (DI) water. After about 45 days, crystals were harvested. The crystals were quickly washed with cold DI water to remove any surface impurities and stored at room temperature. Single crystals for the study were screened using the cross-polarized light microscopy technique~\cite{smallman_characterization_2014} to eliminate twinned and polycrystals; essentially, samples were observed while rotating them under crossed polarizers to verify uniform light transmission.   The single crystals selected for the study had dimensions of approximately $4 \times 5 \times 2\ \text{mm}^3$ for \textit{L-}alanine and $3 \times 5 \times 2\ \text{mm}^3$ for \textit{D-}alanine.  The crystals were glued to a tenon mounting plate for the goniometer NMR probe using epoxy resin.

\subsection{Single-Crystal x-ray Diffraction} \label{section : X-ray}
The lattice structure parameters were determined using powder x-ray diffraction (XRD) analysis at the Electron Microbeam Analysis Laboratory (EMAL), University of Michigan. Powdered \textit{D-} and \textit{L-}alanine samples were analyzed in  reflection mode in Bragg-Brentano geometry on a x-ray diffractometer (Rigaku Ultima IV). The  Cu anode x-ray beam (40 kV, 44 mA) was filtered by a  $20\  \mu m$ thick nickel foil to remove Cu $K\beta$, giving monochromatic Cu $K \alpha$ x-rays. The divergence, scattering and receiving slits were set at $2/3^\circ, \ 2/3^\circ$ and 0.6 mm, respectively. The scanning $2\theta$ range  was from $5^\circ$ to $70^\circ$ with step size of $0.02^\circ$ at a scan rate of $1^\circ\ \text{min}^{-1}$. The unit cell parameters ($a,\ b,\ c,\ \alpha,\ \beta,\ \gamma$) were recovered by assigning the appropriate triple of Miller indices ($h k l$) to each observed inter-planar spacing ($d_{hkl}$).The indexing process was performed using EXPO2014~\cite{altomare_it_2013} via the N-TREOR09 program~\cite{altomare_advances_2009}, the evolution of the N-TREOR software~\cite{altomare_new_2000}. Powder XRD confirmed that the unit cells match the reported crystal structures~\cite{wilson_neutron_2005}; our measured \textit{D-} and \textit{L-}alanine crystal parameters are given in Table~\ref{tab: alanine crystal_data}.

The orientation of the mounted single crystals was determined using the so-called ``Omega-Scan'' method described elsewhere.~\cite{agarwal2025boron}. The mounted crystal's surface normal and edge plane directions were used to find the orientation matrix and ultimately the orientation of the crystal frames relative to the tenon frames. The Euler angles relating the crystal frame to the goniometer frame for the \textit{D-}alanine crystal are $352.7^\circ, 90.0^\circ, 135.0^\circ$ and for the \textit{L-}alanine  crystal are $333.5^\circ, 33.7^\circ, 90.0^\circ$ (see Supplementary Material for details).

\begin{table}[H]
    \centering
    \begin{tabular}{c c c}
    \toprule
    & \textit{L-}alanine & \textit{D-}alanine\\
    \midrule
       Empirical formula  &  \ce{C3H7NO2} & \ce{C3H7NO2} \\
       
       Formula weight  & 75.08 & 75.08\\

       Crystal system & Orthorhombic &  Orthorhombic \\
       
       Space group & $P2_12_12_1$ & $P2_12_12_1$\\
       
        \textit{a} & 6.036 (3) \r{A} & 6.025 (3) \r{A}\\
        \textit{b} & 12.342 (5) \r{A} & 12.324 (5) \r{A}\\
        \textit{c}& 5.788~(3)~\r{A} & 5.783 (3) \r{A}\\
       Z & 4 & 4\\
       
       Cell volume & 431.2 $\text{\AA}^3$ & 429.4 $\text{\AA}^3$\\
       \bottomrule
    \end{tabular}
    \caption{Crystallographic data for \textit{D-}alanine and \textit{L-}alanine at 295 K}
    \label{tab: alanine crystal_data}
\end{table}

\subsection{Single-crystal NMR Spectroscopy}

\begin{figure}[]
\centering
\begin{subfigure}[t]{0.58\textwidth}
        \centering
        \caption{}
        \includegraphics[width=\textwidth]{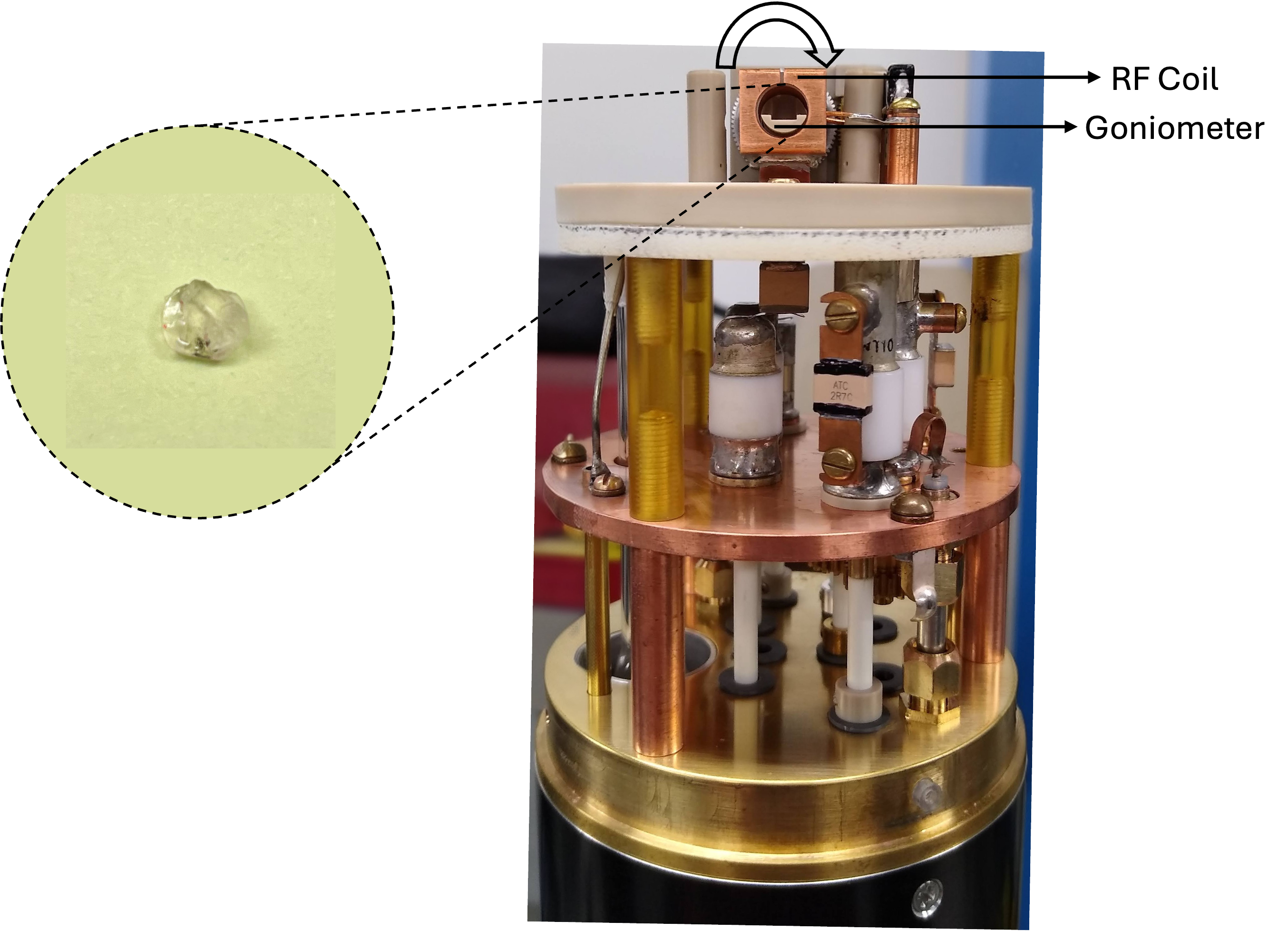}
        
        \label{fig:a_probe}
    \end{subfigure}
    \hfill
    \begin{subfigure}[t]{0.38\textwidth}
        \centering
        \caption{}
        \includegraphics[width=\textwidth]{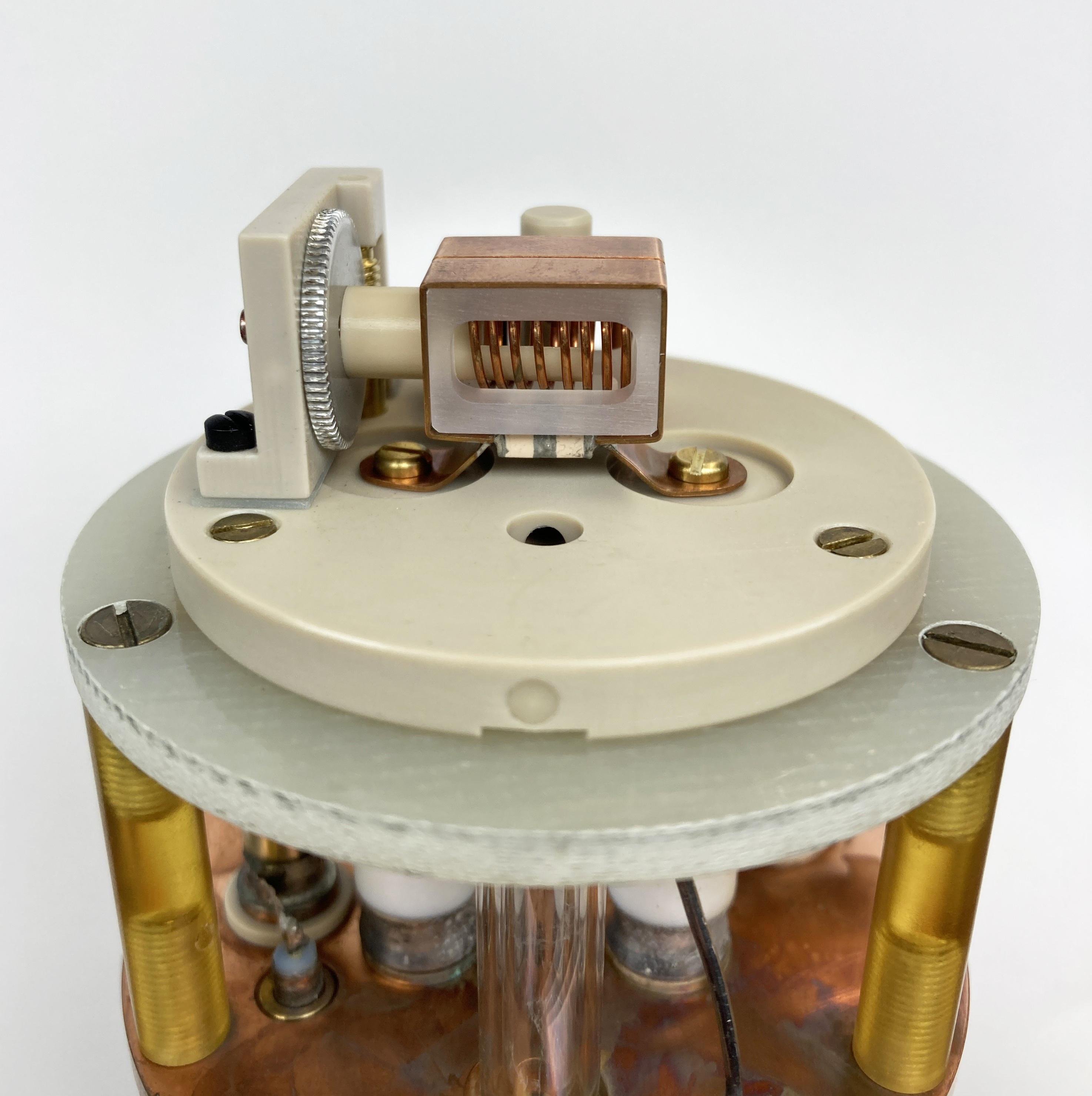}
        
        \label{fig:b_probe}
    \end{subfigure}

    \vspace{2mm}
    \begin{subfigure}[t]{\textwidth}
        \centering
        \caption{}
        \includegraphics[width=\textwidth]{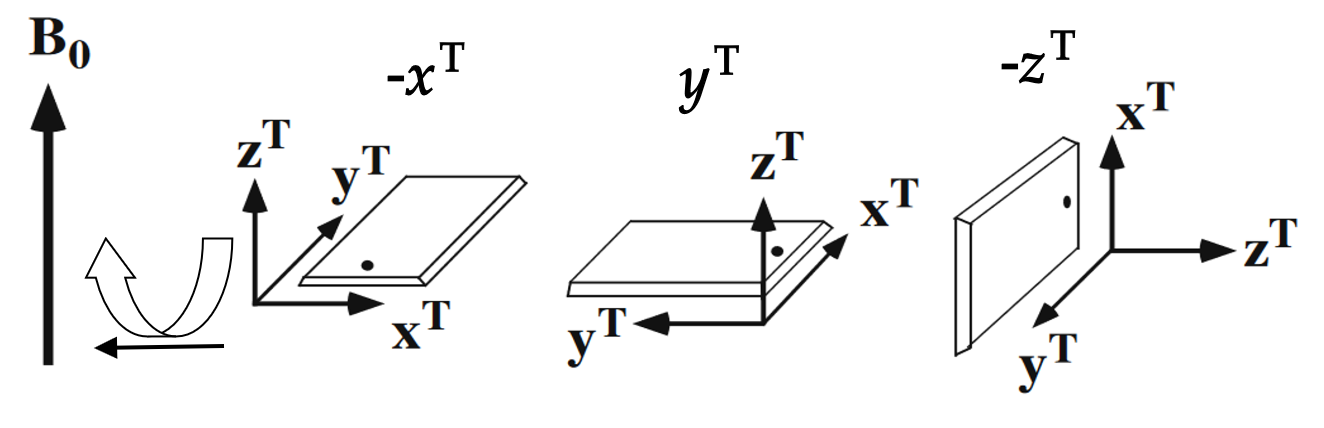}
        
        \label{fig:c_rotation}
    \end{subfigure}
    \caption{\textbf{(a)}: Photograph of front view of the probe. The arrow shows rotation of the goniometer inside the RF coil. Inset shows \textit{L-}alanine crystal glued to the tenon. The tenon may be mounted on three different orientations in the goniometer so that three orthogonal rotations of the sample can be achieved. \\
    \textbf{(b)}: Photograph of side view of the probe showing goniometer mechanism and probe coils.\\
    \textbf{(c)}: Rotation of tenon around an axis that is perpendicular to magnetic field. The three mutually perpendicular rotations are achieved by  mounting the tenon plate into dovetails for $\text{-}x^T$, $y^T$, and $\text{-}z^T$ rotations.  (This Figure is modified from Ref.~\cite{vosegaard_single-crystal_2021} under CC BY 4.0 license)
   }
    \label{fig:probe}
\end{figure}

NMR spectra for both samples were acquired using a Bruker Avance III console running Topspin 3.6 (Bruker Biospin GmbH) at the National High Magnetic Field Laboratory (NHMFL) in Tallahassee, FL. A custom-built low-E 600 MHz static HX probe (Figure~\ref{fig:a_probe} and~\ref{fig:b_probe}), developed at NHMFL, was used for the measurements. The probe features a cross-coil arrangement of a 6.5 mm ID round 9-turn X-channel detection solenoid coil mounted inside and orthogonal to a low inductance \isotope[1]{H}-channel loop gap resonator was employed for the measurements. The low-E coil and associated probe circuitry have been thoroughly described in the literature.~\cite{gorkov_using_2007}

This probe was optimized for \isotope[17]O detection with \isotope[1]H decoupling and operated inside a 600 MHz (14.1 T), 89 mm bore magnet. The tenon plate, with the  crystal glued on it, was mounted in the dovetail track of the goniometer and positioned within the 6 mm inner diameter loop-gap resonator-type NMR sample coil, enabling stepwise rotation to acquire single-crystal NMR spectra across defined rotation patterns. The mounting configuration was designed to allow positive rotations of the tenon about the -$x^T,\ y^T,\text{and}\ \text{-}z^T$ axes, as defined by the dovetail pattern inscribed in the goniometer~\cite{vosegaard_new_1996}. Details of the goniometer mechanism and sample mount construction have been previously described~\cite{agarwal2025boron}. Figure~\ref{fig:probe} shows the probe, the single crystal sample mounted on the tenon, and the three orthogonal tenon rotations relative to the magnetic field, which vary depending on how the plate is mounted in the goniometer.

The mounted samples were rotated from $0^\circ\ \text{to}\ 180^\circ$ using the goniometer's worm-gear mechanism, measured by adjusting an analog micrometer scale at the bottom of the probe. To minimize backlash error, all rotations were carried out in a single direction. The 90-degree pulse lengths were $4\ \mu s$ for the \isotope[17]{O} channel and $2.5\ \mu s$ for the \isotope[1]{H} channel, respectively.  Both the low frequency (\isotope[17]O) and high frequency (\isotope[1]H) channels were manually re-tuned and re-matched at least every other angular increment. All spectra were acquired at room temperature (ca. $22^\circ$C) with air cooling. 

For each rotation axis, spectra were recorded with spectral width of 200 kHz and acquisition time of 0.02 s for \textit{L-}alanine and spectral width of 100 kHz and acquisition time of 0.04 s for \textit{D-}alanine. Between 256 and 512 transients were collected per spectrum, as needed to maintain a consistent signal-to-noise ratio (SNR). The  rotation patterns were verified by observing smooth curves connecting the recorded resonance frequencies and by comparing the spectra at the  starting and ending rotational positions, as well as key check points such as $0^\circ (\text{-}x^T) = 0^\circ(y^T),\   90^\circ (\text{-}x^T) = 90^\circ(\text{-}z^T),\ \text{and}\ 0^\circ (\text{-}z^T) = 90^\circ(y^T)$. Spectral calibration was carried out using an external reference sample of tap water with the \isotope[17]{O} chemical shift defined at 0.0 ppm.

\subsection{Computational Modeling}

The literature neutron-diffraction crystal structures~\cite{wilson_neutron_2005} were used as the bases for the corresponding computed structures for \textit{L-}alanine at 60 K (278467.cif) and 295 K (278464.cif) and for \textit{D-}alanine at 60 K (278466.cif).  The reference structure reported to be \textit{D-}alanine at 295 K (278465.cif), was found to be the \textit{L-}enantiomer, so a proxy for the \textit{D-}enantiomer was created by inverting the \textit{c}-axis of the \textit{L-}alanine.
The 60 K \textit{D-}alanine structure has the N-C*-C angle oriented opposite those in the \textit{L-}alanine structures, relative to the \textit{a}-axis, so the literature structure was transformed by rotation about the \textit{c}-axis to obtain equivalent computational starting points, where the only major differences between the two enantiomer structures was the relative orientation of the hydrogen and methyl groups on the anomeric carbons (C*). All crystal operations were carried out using VESTA~\cite{vesta}. Conversion of the neutron structure .cif files to the CASTEP .cell input files was completes using cif2cell~\cite{cif2cell}.

Although it is common practice to optimize experimental crystal structures prior to calculating magnetic parameters, we also calculated the electric field gradient (EFG) and chemical shielding tensors for the unoptimized structures to enable direct comparison with a previous literature report~\cite{veinberg201614n}. These calculations were performed using both the rPBE functional~\cite{RPBE} with a 650 eV cutoff and the PBEsol functional with a 741 eV cutoff employing a 2 1 2 Monkhorst-Pack k-point grid~\cite{MPgrid} in both cases (see Supplementary Material).

The Electric Field Gradients~\cite{Elec_Field_grad} (EFG) and chemical shielding tensors~\cite{Mag_shielding_tens} of both crystal enantiomers were computed through ultra-soft pseudopotentials~\cite{Ultrasoft_Pseudo_Pot, J_Coupling_usp} using GIPAW (Gauge Including Projector Augmented Waves) approach~\cite{GIPAW, magres_review}, as implemented in the CASTEP-NMR package. The NMR parameters were then calculated using Eqs.~\ref{eq: CS iso} -~\ref{eq: CS order} and Eqs.~\ref{eq: delta Quadrupolar} -~\ref{eq: CQ order} and the mutual orientation of the two was tensors determined using Eq.~\ref{eq: Euler orientation}.

For comparison, the EFG and shielding tensors were also calculated using optimized neutron structures (see Supplementary Material).  These calculations were conducted under 3D periodic boundary conditions using the CASTEP software package~\cite{CASTEP}, employing the PBEsol exchange-correlation functional~\cite{ hohenberg1964inhomogeneous,DFT-KS,perdew2008restoring}, and “precise” basis set precision with automatic finite-basis-set correction~\cite{FBSC}. The planewave cutoff energy was 741 eV, and a 2 1 2 Monkhorst-Pack k-point grid was again used. During optimization, unit-cell parameters were fixed to their literature values, while atomic positions were allowed to relax~\cite{RMP-Payne, BFGS, LBFGS}. The minor differences between the experimental and optimized structures are summarized as root-mean-square displacements in Table~\ref{tab:RMS_Changes}.

\begin{table}[]
    \centering
    \begin{tabular}{c c c c c c}
        \toprule
         & & \multicolumn{4}{c}{RMS deviation / $10^{-4}$ \AA} \\
         \cmidrule(lr){3-6}
         Nucleus & \# centers & \textit{L-}ala (295 K) & \textit{D-}ala (295 K)\textsuperscript{$\ast$} & \textit{L-}ala (60 K) & \textit{D-}ala (60 K) \\
        \midrule
            H & 28 & 2.48 & 2.50 & 1.72 & 1.10 \\
            C & 12 & 5.53 & 5.60 & 3.21 & 5.16 \\
            N & 4 & 1.54 & 1.47 & 1.59 & 0.63 \\
            O & 8 & 9.23 & 7.70 & 9.73 & 9.36 \\
            \hline
            All & 52 & 4.87 & 4.46 & 4.33 & 4.51 \\[0.1ex]
        \bottomrule
\end{tabular}

\begin{minipage}{\linewidth}
\footnotesize
    $^\ast$ Starting structure derived from \textit{L-}alanine (295 K).
    \end{minipage}
    \caption{The average root-mean-square deviations in nuclear positions after optimization of \textit{L-} and \textit{D-}alanine from literature crystal structures.\\}
    \label{tab:RMS_Changes}
\end{table}

\begin{table}[H]  
    \centering
     \begin{adjustbox}{width=\columnwidth,center}
    \begin{tabular}{c c c}
        \toprule
          Component & \multicolumn{2}{c}{Enantiomer} \\
         \cmidrule(lr){2-3} 
         \\
         & \textit{L-}alanine & \textit{D-}alanine \\
       \midrule 
        \\
        Quadrupolar &
        
        $\begin{pmatrix}
 -4.120  & 0.726& -1.187\\
  0.726& -3.623&  1.702\\
 -1.187&  1.702&  7.743\\  
        \end{pmatrix}$ 
        & 
        $\begin{pmatrix}
   -4.120 &  0.726 & 1.187\\
   0.726& -3.623& -1.702\\
   1.187& -1.702&  7.743 \\ 
        \end{pmatrix}$ 
        \\
        \\
        Total CS &
        $\begin{pmatrix}
    -51.128&  209.778&   55.794\\
   180.674&  120.849 & -83.964\\
    19.534&  -46.887& -161.699 \\   
        \end{pmatrix}$
        &
        $\begin{pmatrix}
    -51.139&  209.781&  -55.794\\
   180.676& 120.751 &  83.968\\
   -19.532 &  46.886& -161.799 \\  
        \end{pmatrix}$
        \\
        \\
        Isotropic CS &
        $\begin{pmatrix}
   -30.659  & 0   &   0  \\  
    0 &  -30.659 &  0 \\   
    0  &    0  &  -30.659   \\
        \end{pmatrix}$
        &
        $\begin{pmatrix}
   -30.729 &  0    &  0 \\   
    0  &  -30.729 &  0 \\   
    0  &    0   & -30.729  \\  
        \end{pmatrix}$
        \\
        \\        
        Symmetric CS &
        $\begin{pmatrix}
   -51.128 & 195.226  & 37.664\\
   195.226 & 120.849 & -65.425\\
    37.664 & -65.425& -161.699 \\   
        \end{pmatrix}$
        &
        $\begin{pmatrix}
    -51.139  &195.229 & -37.663\\
   195.229 & 120.751 &  65.427\\
   -37.663  & 65.427& -161.799 \\ 
        \end{pmatrix}$
        \\
        \\
        Antisymmetric CS &
        $\begin{pmatrix}
     0  &   14.552 & 18.130  \\
 -14.552 &  0   & -18.539\\
 -18.130 &  18.539 &  0  \\          
        \end{pmatrix}$
        &
        $\begin{pmatrix}
     0  &   14.553& -18.131\\
 -14.553 &  0 &   18.541\\
   18.131 &-18.541&   0  \\          
        \end{pmatrix}$        
        \\
        \\
        \bottomrule
    \end{tabular}
    \end{adjustbox}
        \caption{Calculated quadrupolar and chemical shielding (CS) tensors (rounded to 3 decimal places) for one of the magnetically equivalent \isotope [17] {O} site for crystallographic site O1 in enantiomers of alanine.}
    \label{tab:17O site1 tensors}
\end{table}

\begin{table}[H]  
    \centering
     \begin{adjustbox}{width=\columnwidth,center}
    \begin{tabular}{c c c}
        \toprule
          Component & \multicolumn{2}{c}{Enantiomer} \\
         \cmidrule(lr){2-3} 
         \\
         & \textit{L-}alanine & \textit{D-}alanine \\
       \midrule 
        \\
Quadrupolar &
        
        $\begin{pmatrix}
    0.942& -1.012 & 4.634\\
 -1.012 &-0.817& -3.580  \\
   4.634 &-3.580 & -0.125 \\
        \end{pmatrix}$ 
        & 
        $\begin{pmatrix}
    0.942& -1.012& -4.634\\
 -1.012& -0.817 & 3.580  \\
 -4.634 & 3.580 & -0.125 \\
        \end{pmatrix}$ 
        \\
        \\
        Total CS &
        $\begin{pmatrix}
   -63.835 &165.365 &-76.405\\
 168.444 & 77.812 & 26.025\\
    7.265 &-23.272 &-49.805 \\  
        \end{pmatrix}$
        &
        $\begin{pmatrix}
   -63.854& 165.377&  76.405\\
 168.449 & 77.714& -26.028\\
   -7.269 & 23.276 &-49.924  \\ 
        \end{pmatrix}$
        \\
        \\
        Isotropic CS &
        $\begin{pmatrix}
   -11.943 &  0    &  0    \\
    0  &  -11.943 &  0    \\
    0  &    0   & -11.943 \\
        \end{pmatrix}$
        &
        $\begin{pmatrix}
   -12.021 &  0   &   0   \\ 
    0  &  -12.021  & 0    \\
    0   &   0  &  -12.021  \\
        \end{pmatrix}$
        \\
        \\        
        Symmetric CS &
        $\begin{pmatrix}
   -63.835& 166.905& -34.57  \\
 166.905&  77.812 &  1.376\\
 -34.570&   1.376& -49.805  \\  
        \end{pmatrix}$
        &
        $\begin{pmatrix}
 -63.854& 166.913&  34.568\\
 166.913 & 77.714&  -1.376\\
  34.568 & -1.376& -49.924 \\
        \end{pmatrix}$
        \\
        \\
        Antisymmetric CS &
        $\begin{pmatrix}
   0  &   -1.539& -41.835\\
   1.539 &  0  &   24.648\\
  41.835& -24.648 &  0     \\
        \end{pmatrix}$
        &
        $\begin{pmatrix}
   0  &   -1.536 & 41.837\\
   1.536 &  0 &   -24.652\\
 -41.837 & 24.652 &  0      \\
        \end{pmatrix}$         
        \\
        \\
        \bottomrule
    \end{tabular}
    \end{adjustbox}
        \caption{Calculated quadrupolar and chemical shielding (CS) tensors (rounded to 3 decimal places) for one of the magnetically equivalent \isotope [17] {O} site for crystallographic site O2 in enantiomers of alanine.}
    \label{tab:17O site2 tensors}
\end{table}

\begin{table}[H]  
    \centering
     \begin{adjustbox}{width=\columnwidth,center}
    \begin{tabular}{c c c}
        \toprule
          Component & \multicolumn{2}{c}{Enantiomer} \\
         \cmidrule(lr){2-3} 
         \\
         & \textit{L-}alanine & \textit{D-}alanine \\
       \midrule 
        \\
 Quadrupolar &
        
        $\begin{pmatrix}
  0.433& -0.236& -0.975\\
 -0.236& -0.421&  0.283\\
 -0.975&  0.283& -0.012
        \end{pmatrix}$ 
        & 
        $\begin{pmatrix}
 0.433& -0.236 & 0.975\\
 -0.236& -0.421 &-0.283\\
  0.975& -0.283& -0.012
        \end{pmatrix}$ 
        \\
        \\
        Total CS &
        $\begin{pmatrix}
 188.770  &  1.781 & -4.136\\
 -1.286& 176.790  & -2.976\\
  -4.771 & -3.925& 193.611

        \end{pmatrix}$
        &
        $\begin{pmatrix}
 188.766 &  1.779  & 4.136\\
  -1.286& 176.696 &  2.980 \\
   4.770 &   3.925& 193.518
        \end{pmatrix}$
        \\
        \\
        Isotropic CS &
        $\begin{pmatrix}
 186.390  & 0  &   0  \\
   0 &  186.390  & 0  \\
  0  &   0 &  186.390
        \end{pmatrix}$
        &
        $\begin{pmatrix}
 186.327 &  0   &   0   \\
  0  &  186.327 &  0  \\
   0  &   0   & 186.327
        \end{pmatrix}$
        \\
        \\        
        Symmetric CS &
        $\begin{pmatrix}
 188.770  &  0.248 & -4.453\\
   0.248& 176.790 &  -3.450 \\
  -4.453 & -3.450 & 193.611 
        \end{pmatrix}$
        &
        $\begin{pmatrix}
 188.766 &  0.246 &  4.453\\
   0.246& 176.696 &  3.453\\
  4.453 &  3.453& 193.518 
        \end{pmatrix}$
        \\
        \\
        Antisymmetric CS &
        $\begin{pmatrix}
  0  &   1.534 & 0.318\\
 -1.534&  0  &   0.474\\
 -0.318& -0.474 & 0          
        \end{pmatrix}$
        &
        $\begin{pmatrix}
  0  &   1.532& -0.317\\
 -1.532 & 0 &   -0.472\\
  0.317 & 0.472 & 0     
        \end{pmatrix}$         
        \\
        \\
        \bottomrule
    \end{tabular}
    \end{adjustbox}
        \caption{Calculated quadrupolar and chemical shielding (CS) tensors (rounded to 3 decimal places) for one of the magnetically equivalent \isotope [14] {N} site in enantiomers of alanine.}
    \label{tab:14N site tensors}
\end{table}


\section{Results and discussion}

\begin{figure}[H]
    \centering
    \includegraphics[width=\textwidth]{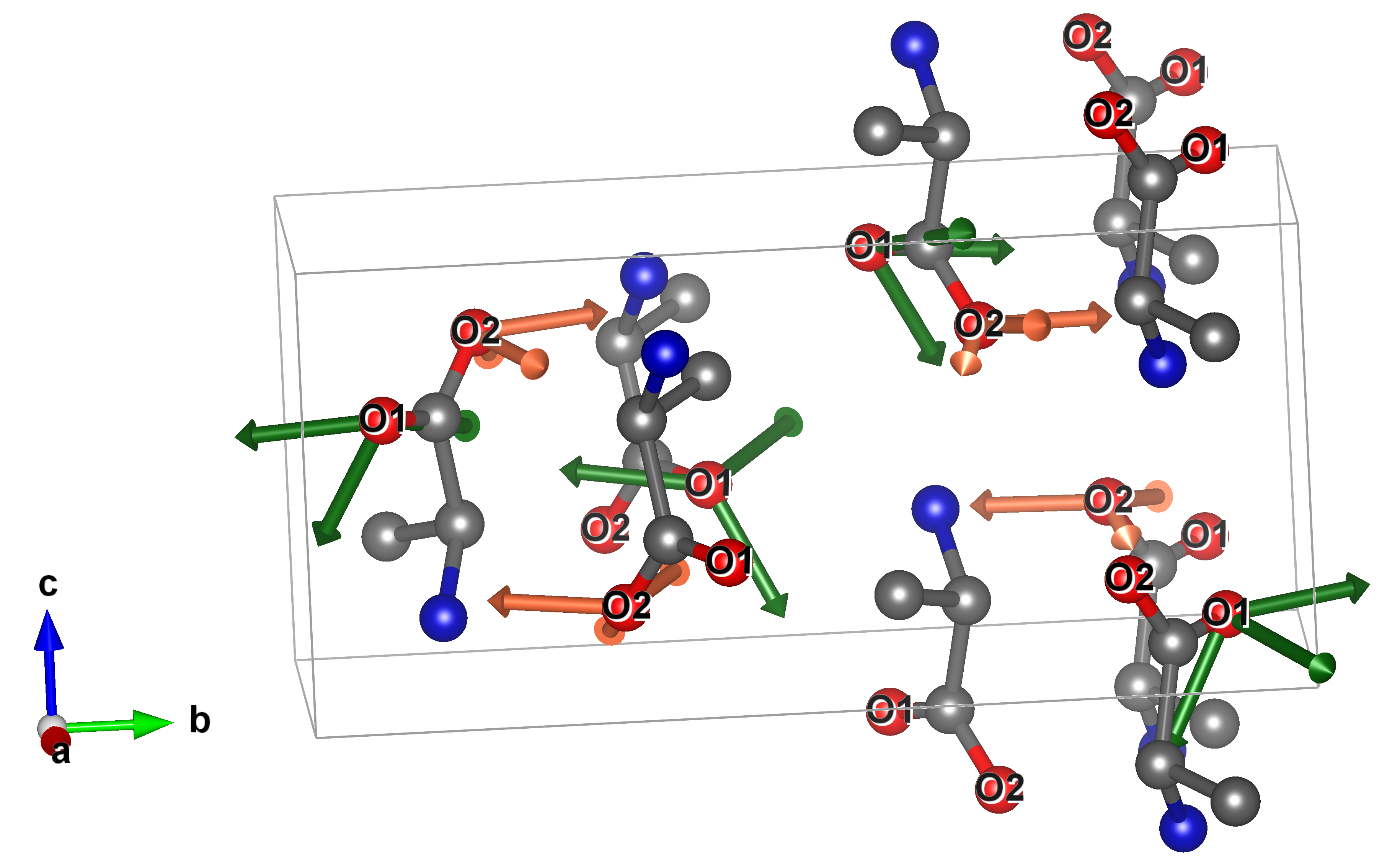}
 \caption{Orthorhombic crystal structure of alanine  with crystallographically nonequivalent O nuclei labeled as O1 and O2 and their shielding tensors represented in orange and green colors. All the eight O nuclei are magnetically nonequivalent. The arrows represent directions and scaled magnitudes of the computed chemical shielding tensors in the crystal frame.  O, C, and N nuclei are shown in red, gray, and blue, respectively.   (H nuclei omitted for clarity.)}
    \label{fig:unit cell alanine}
\end{figure}

As shown in Figure~\ref{fig:unit cell alanine}, the unit cell for alanine has two  nonequivalent crystallographic  O sites with four magnetically nonequivalent O nuclei in each site related by screw symmetry. As a result, an independent resonance frequency is expected  from each of the magnetically nonequivalent O nuclei. Also shown in the figure are the calculated directions and scaled magnitudes of the O-center chemical shielding tensors.

\begin{figure}[]
    \centering
        \begin{subfigure}{0.32\textwidth}
        \caption{}
        \includegraphics[width=1\linewidth]{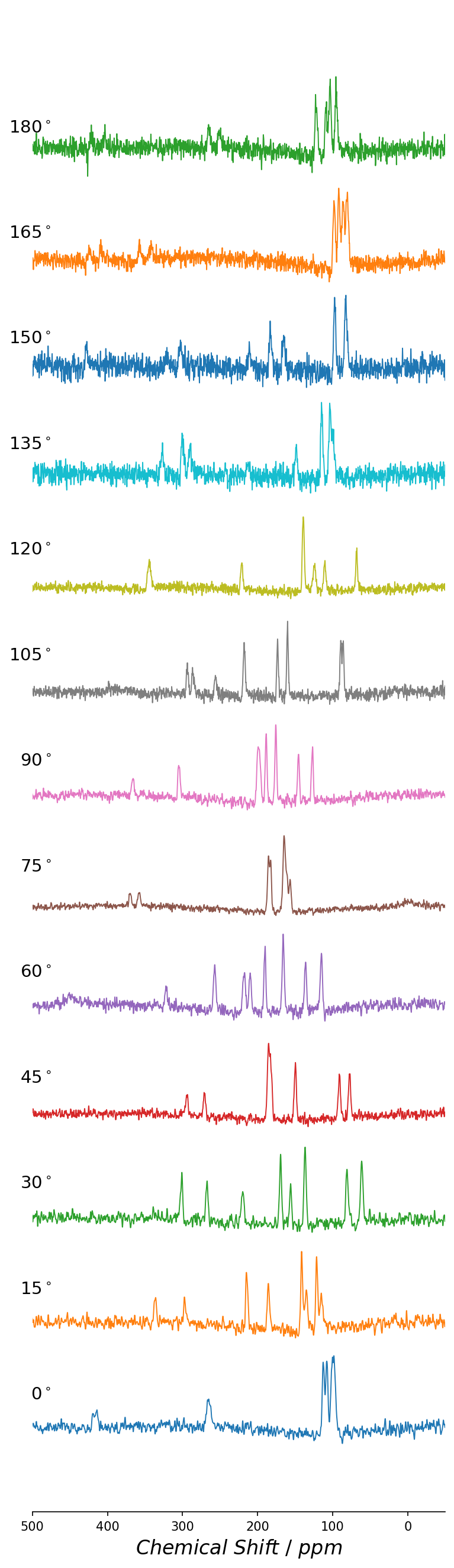}
        \end{subfigure}
        \begin{subfigure}{0.32\textwidth}
        \caption{}
        \includegraphics[width=1\linewidth]{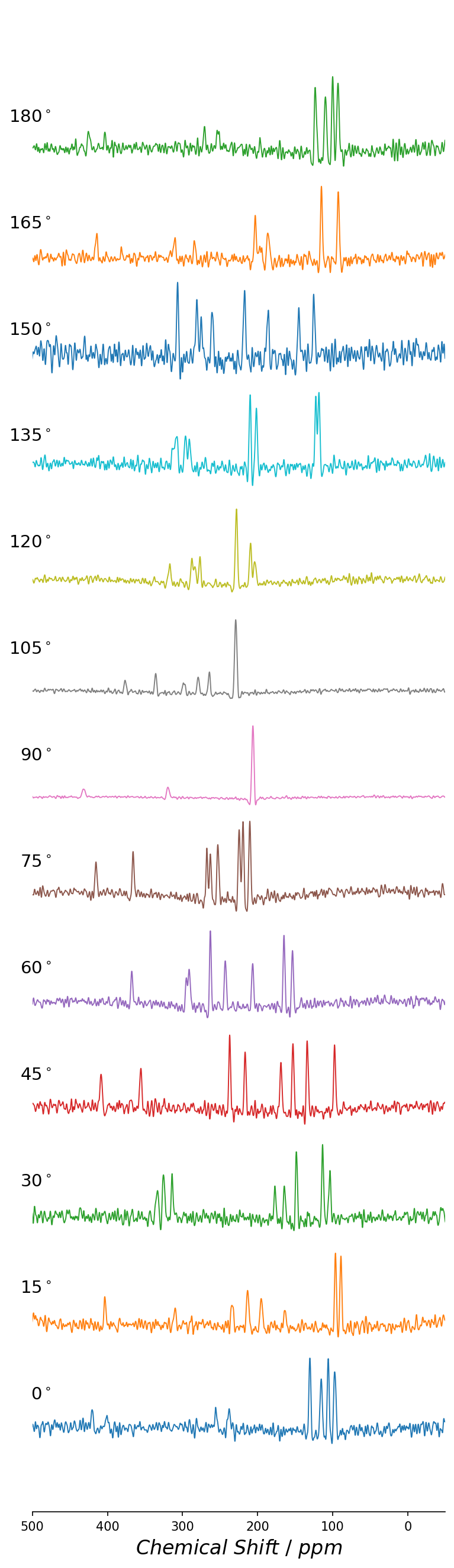}
        \end{subfigure}
        \begin{subfigure}{0.32\textwidth}
        \caption{}
        \includegraphics[width=1\linewidth]{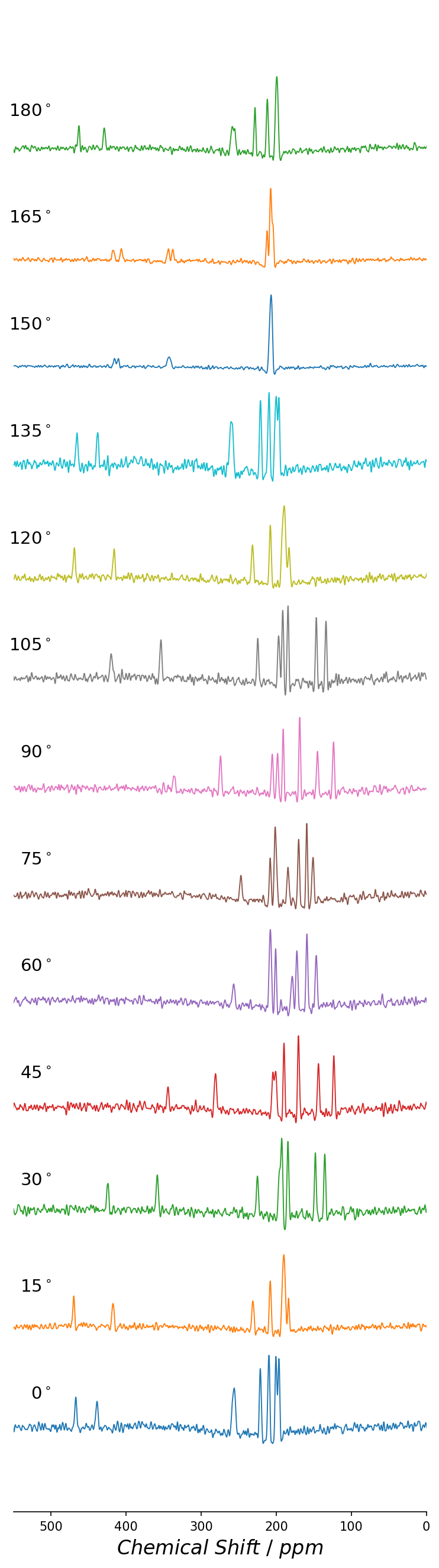}
        \end{subfigure}
    \caption{Single-crystal \isotope[17]{O} NMR spectra at 14.1 T for \textit{L-}alanine. The spectra were recorded at $15^\circ$ increments for rotation about \textbf{(a)} mounting -$x^T$, \textbf{(b)} mounting $y^T$, and \textbf{(c)} mounting -$z^T$
    }
    \label{fig:stacked plot Lalanine}
\end{figure}

\begin{figure}[]
    \centering
        \begin{subfigure}{0.32\textwidth}
        \caption{}
        \includegraphics[width=1\linewidth]{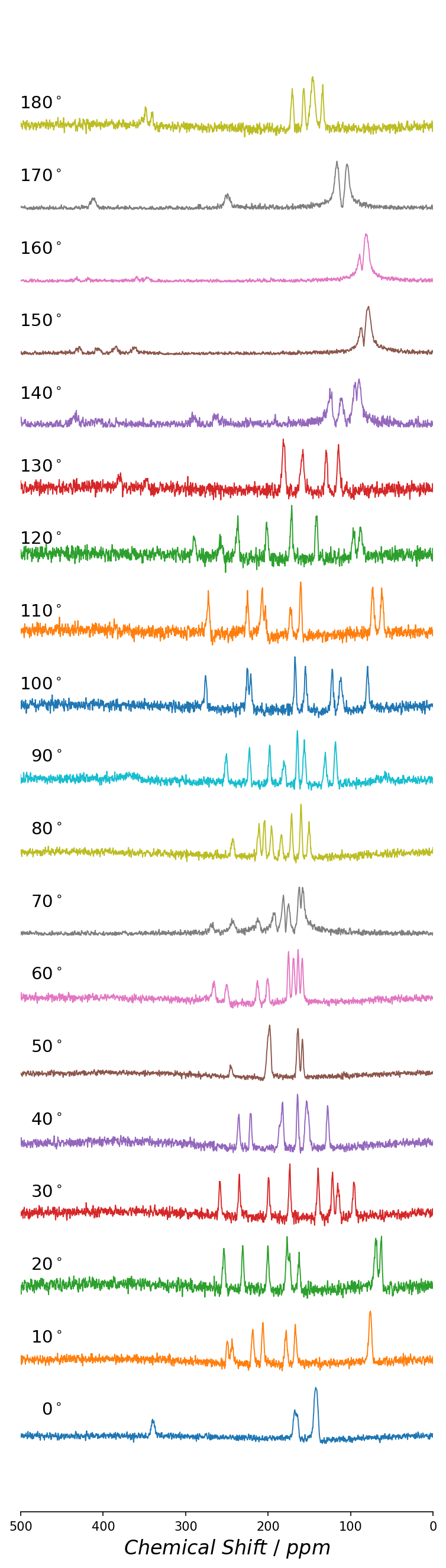}
        \end{subfigure}
        \begin{subfigure}{0.32\textwidth}
        \caption{}
        \includegraphics[width=1\linewidth]{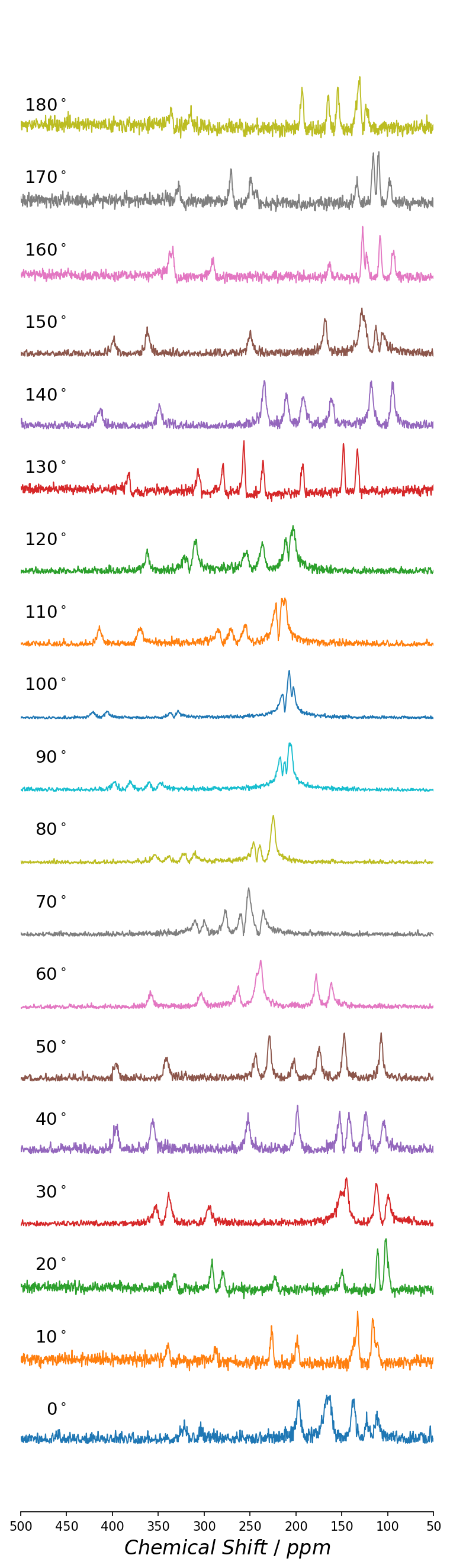}
        \end{subfigure}
        \begin{subfigure}{0.32\textwidth}
        \caption{}
        \includegraphics[width=1\linewidth]{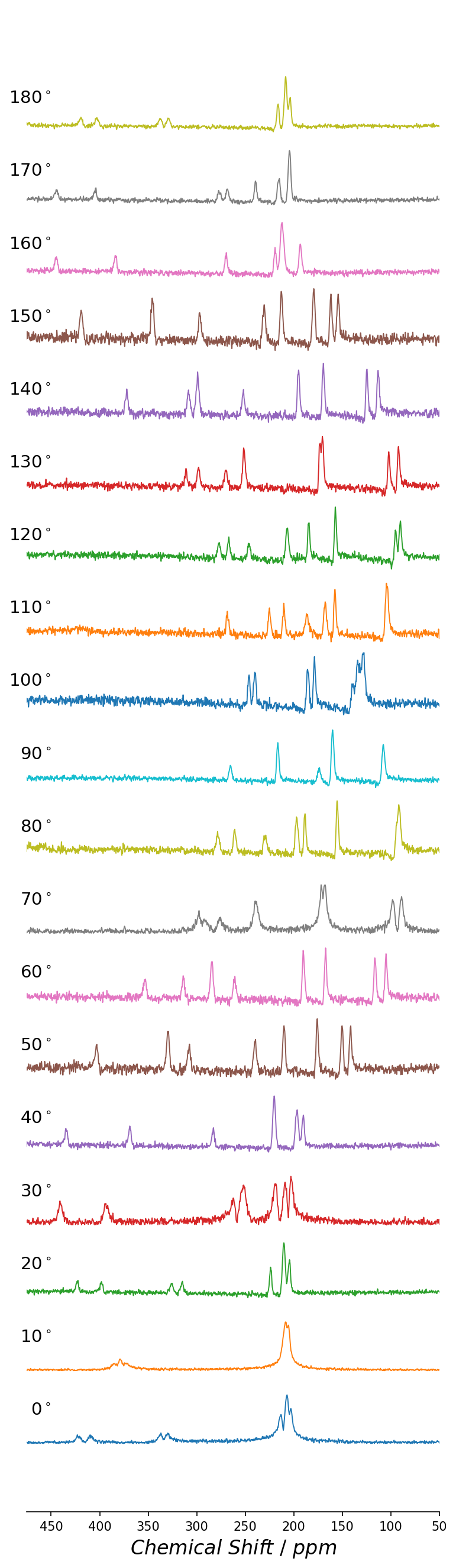}
        \end{subfigure}
    \caption{Single-crystal \isotope[17]{O} NMR spectra at 14.1 T for \textit{D-}alanine. The spectra were recorded at $10^\circ$ increments for rotation about \textbf{(a)} mounting -$x^T$, \textbf{(b)} mounting $y^T$, and \textbf{(c)} mounting -$z^T$
    }
    \label{fig:stacked plot Dalanine}
\end{figure}

The single-crystal \isotope[17]{O} NMR spectra for \textit{L-}alanine rotation about -$x^T$, $y^T$, and -$z^T$ axes are shown in Figure~\ref{fig:stacked plot Lalanine} and those for \textit{D-}alanine in Figure~\ref{fig:stacked plot Dalanine}. A maximum of eight equally intense transitions for each orientation of the single crystal were observed, as expected from eight magnetically nonequivalent O sites. While we were able to observe the satellite transitions (see Supplementary Material), limited time and resources prevented us from measuring the systematic shift of satellite transitions required to extract the quadrupolar ACS cross-coupling terms.

The correlations of the rotation plots and peak assignments was assisted by the fact that the NMR chemical shift frequency remains the same for certain pairs of  crystal orientations 
\begin{align}
    \text{Mounting -} x^T\   (\Theta = 0^\circ)\  &\equiv  \text{Mounting } y^T\ (\Theta = 0^\circ)\label{eq:mounting1}\\
       \text{Mounting -} x^T\   (\Theta = 90^\circ)\  &\equiv  \text{Mounting  -}z^T\ (\Theta = 90^\circ)\label{eq:mounting2}\\
            \text{Mounting }y^T\  (\Theta = 90^\circ)\  &\equiv  \text{Mounting -}z^T\ (\Theta = 0^\circ)\label{eq:mounting3}
\end{align}
where the mountings are as shown in Figure~\ref{fig:c_rotation}.
This pairwise coincidence results from NMR
interactions being insensitive to rotation parallel to the magnetic field
axis~\cite{vosegaard_improved_1998}.


\subsection{Analysis of Single-Crystal NMR Spectra} \label{section: alanine analysis}

\begin{figure}[H]
    \centering
    \includegraphics[width=1\linewidth]{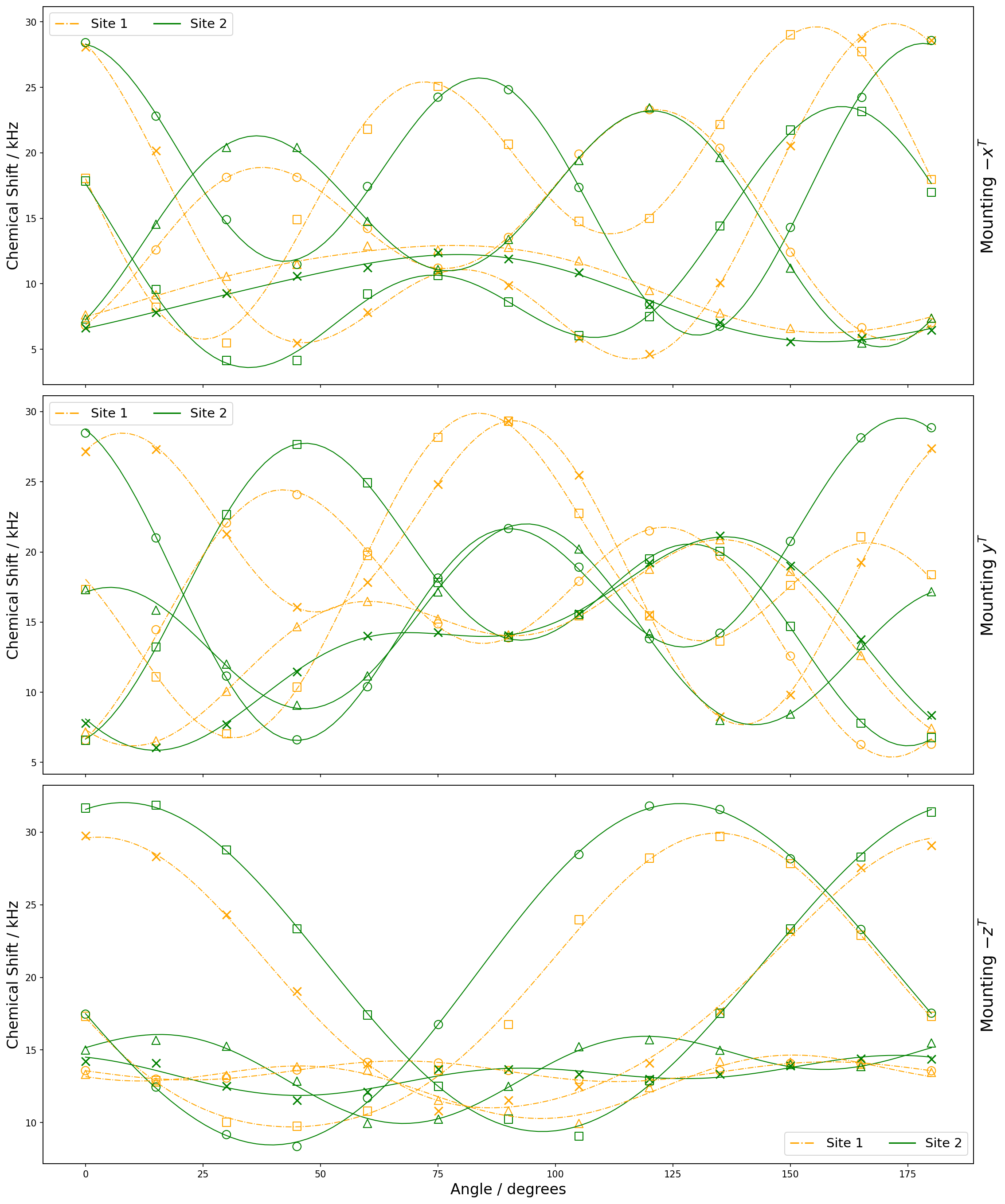}
    \caption{Rotation plots for \isotope[17]{O} central transition in \textit{L-}alanine showing experimental resonances with two crystallographically nonequivalent \isotope[17]{O} sites each having four magnetically nonequivalent \isotope[17]{O} nuclei (marked as $\bigcirc, \square, \bigtriangleup, \cross$) under rotation of the three orientations of the crystal sample. The curves are constructed from optimized coefficients (see Table~\ref{tab:Lala fit coeff}).}
    \label{fig:Lalanine rotation plot}
\end{figure}

\begin{figure}[H]
    \centering
    \includegraphics[width=1\linewidth]{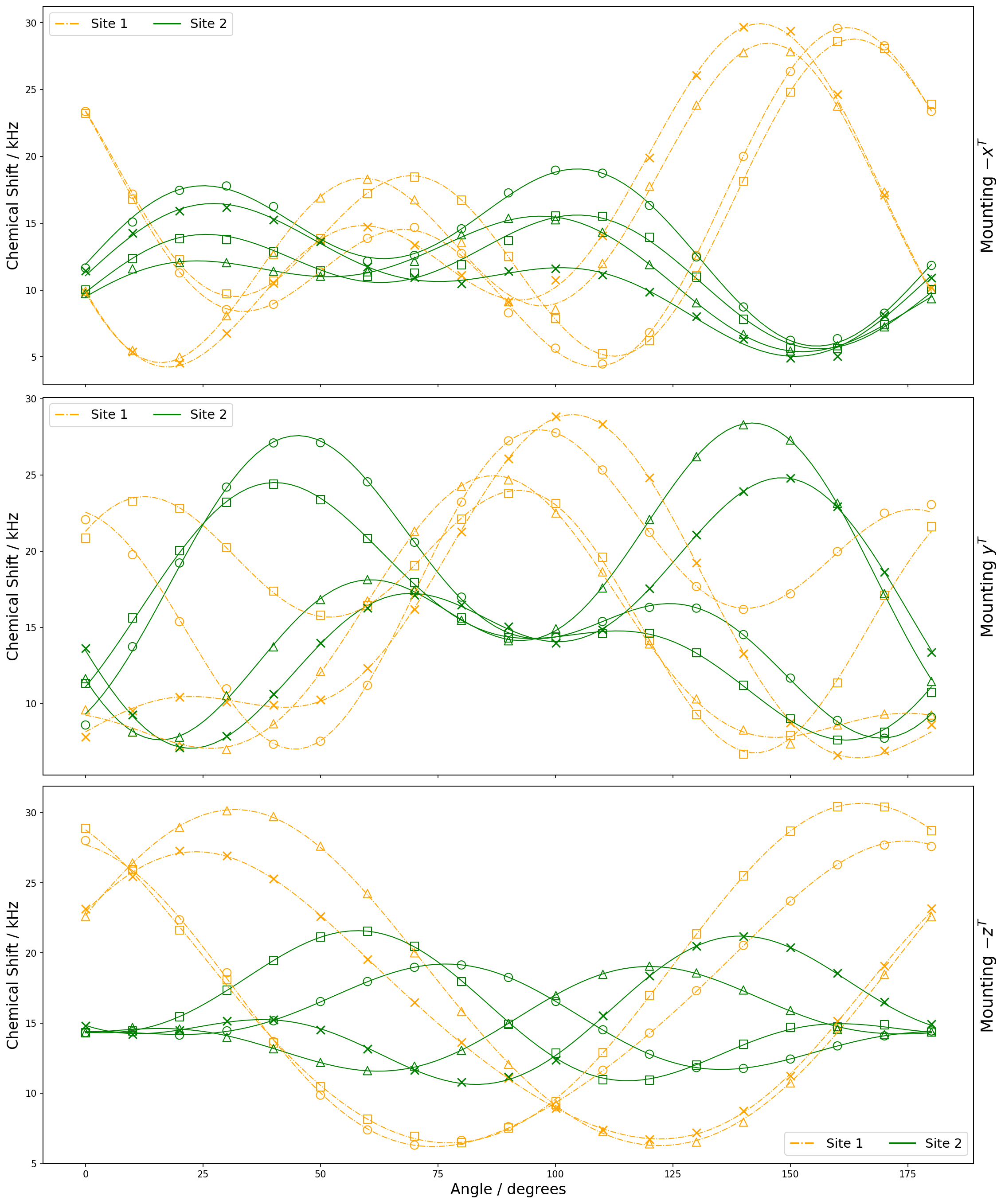}
    \caption{Rotation plots for \isotope[17]{O} central transition in \textit{D-}alanine showing experimental resonances with two crystallographically nonequivalent \isotope[17]{O} sites each having four magnetically nonequivalent \isotope[17]{O} nuclei (marked as $\bigcirc, \square, \bigtriangleup, \cross$) under rotation of the three orientations of the crystal sample. The curves are constructed from optimized coefficients (see Table~\ref{tab:Dala fit coeff}).}
    \label{fig:Dalanine rotation plot}
\end{figure}

The optimized quadrupolar and CSA parameters, along with their errors, were obtained by fitting the central transition of each site using the  Analysis of Single-Crystal Spectra (ASICS) software package~\cite{vosegaard_improved_1998}. The rotation plots shown in Figures~\ref{fig:Lalanine rotation plot} and~\ref{fig:Dalanine rotation plot} were fit according to Eqs.~\ref{eq: CS fit equation} -~\ref{eq: Q fit equation}. Tables~\ref{tab:Lala fit coeff} and~\ref{tab:Dala fit coeff} summarize the optimized coefficients provided by ASICS for magnetically nonequivalent O nuclei in  \textit{L-}alanine and \textit{D-}alanine.

\begin{table}[H]
    \centering
\begin{tabular}{c c c c c c c}
\toprule
 Mounting & Nucleus & $A$ & $B$ & $C$ & $D$ & $E$\\
 \midrule
    -$x^T$ & 1 & 14.7& -3.37& -1.05& -4.51& 4.25\\
    $y^T$ &  & 20.2& -1.06& 3.90& 8.08& 2.59\\
    -$z^T$ &  & 20.6& 10.9& 2.99& 0.0857& 0.151 \\[1ex]
    -$x^T$ & 2 & 15.1& -3.05& 0.266& -4.81& 5.09\\
    $y^T$ &  & 17.8& 3.54& -3.81& 7.41& -1.81 \\
    -$z^T$ &  & 19.4& 9.26& 0.309 &0.972 &0.523\\[1ex]
    -$x^T$ & 3 & 17.9& 1.69& 2.62 &8.68 &-2.39\\
    $y^T$ &  & 13.7& -2.99& -4.76& -2.61& -1.80 \\
    -$z^T$ &  & 18.6& -0.238&-10.1& -1.17& 0.00316\\[1ex]
    -$x^T$ & 4 & 11.4 &4.52 &-4.75 &1.77 &-5.42\\
    $y^T$ &  & 14.2& -3.35& -3.13& -3.54& -1.75 \\
    -$z^T$ &  & 20.2& -2.81& -11.4& 0.125& 0.289\\[1ex]
    -$x^T$ & 5 & 18.6 &-1.19 &-4.39 &0.646 &-8.69\\
    $y^T$ &  & 17.0&-3.58& 3.85& -6.80& 0.814 \\
    -$z^T$ &  & 13.8& 1.36& -1.27& 0.00303& 2.00\\[1ex]
    -$x^T$ & 6 & 13.5 &9.16 &-2.22 &5.72 &-4.05\\
    $y^T$ &  & 16.1& -3.60& 2.31& -5.84& 3.21 \\
    -$z^T$ &  & 13.3& 0.387& -0.628& 0.812& -0.167\\[1ex]
    -$x^T$ & 7 & 9.90 &-2.60&2.00 &0.187 &0.391\\
    $y^T$ &  & 17.8& -5.53& -1.83& 5.76& -4.37 \\
    -$z^T$ &  & 13.5& 0.0178& 0.0635& 0.00522& -0.667\\[1ex]
    -$x^T$ & 8 & 8.95 &-2.69 &1.82 &0.333 &0.188\\
    $y^T$ &  & 14.0 &-2.32&0.210 &5.41& 1.76 \\
    -$z^T$ &  & 12.8& 1.35& 0.0144& -1.04& -0.714\\
    \bottomrule
    
\end{tabular}
    \caption{Optimized coefficients (in kHz) for \isotope[17]{O} rotation data of \textit{L-}alanine (up to 3 significant figures.) }
    \label{tab:Lala fit coeff}
\end{table}

\begin{table}[H]
    \centering
\begin{tabular}{c c c c c c c}
\toprule
 Mounting & Nucleus & $A$ & $B$ & $C$ & $D$ & $E$\\
  \midrule
    -$x^T$ & 1 & 14.7& 7.22& -3.12& 1.49& -7.19\\
    $y^T$ &  & 16.8& -2.66& 5.99& -4.81& 1.26\\
    -$z^T$ &  & 17.7& 5.26& 10.8& -0.297& 0.450 \\[1ex]
    -$x^T$ & 2 & 15.6& 5.46& -1.24& 2.31& -7.60\\
    $y^T$ &  & 15.5& -1.64& 5.97& -2.77& 2.32 \\
    -$z^T$ &  & 16.4& 5.96& 8.18&0.555& 0.773\\[1ex]
    -$x^T$ & 3 & 15.2& 0.0720& -5.50& -5.49& -5.99\\
    $y^T$ &  & 17.3& -1.32& 4.26& 5.32& 2.95 \\
    -$z^T$ &  & 17.9 &10.7 &-5.69& 0.159& -0.638\\[1ex]
    -$x^T$ & 4 & 15.0& 0.311& -8.01& -5.37& -4.87\\
    $y^T$ &  & 18.4& -2.32& -4.82& 6.52& 0.922 \\
    -$z^T$ &  & 16.5 &10.1& -3.54& 1.14& 0.382\\[1ex]
    -$x^T$ & 5 & 13.6& -2.57& 2.12& 0.876& 4.43\\
    $y^T$ &  & 17.2 &-1.31 &-6.04&-4.35&-3.78 \\
    -$z^T$ &  & 15.6& -0.358& 3.83& -0.928& -2.24\\[1ex]
    -$x^T$ & 6 & 11.4& -2.15& 1.44& 0.601& 3.22\\
    $y^T$ &  & 15.8 &-0.694 &-5.21 &-1.64 &-4.58 \\
    -$z^T$ &  & 15.0& -1.94& 2.06& 1.29& -0.925\\[1ex]
    -$x^T$ & 7 & 10.9& -2.86& 1.726& 1.46& 2.08\\
    $y^T$ &  & 15.1 &-8.97& -3.20 &2.06 &4.02 \\
    -$z^T$ &  & 15.0& -0.337& -2.65& -0.337& 1.64\\[1ex]
    -$x^T$ & 8 & 10.9& -0.0933& 3.80& 0.263& 2.75\\
    $y^T$ &  & 13.3& -7.79& 0.598& 3.76& -0.998 \\
    -$z^T$ &  & 15.5& 1.93& -3.02& -2.57& -0.0608\\
    \bottomrule
    
\end{tabular}
    \caption{Optimized coefficients (in kHz) for \isotope[17]{O} rotation data of \textit{D-}alanine (up to 3 significant figures.) }
    \label{tab:Dala fit coeff}
\end{table}

Table~\ref{tab:alaine results} summarizes the optimized experimental parameters, with error limit estimated as 95\% confidence intervals for individual parameters, along with the DFT computations of quadrupolar coupling and CSA parameters. For some parameters, the confidence interval could not be estimated as chi-square values (goodness of fit) for those parameters were not normally distributed and did not converge to a minimum. The two anionic oxygen sites in the zwitterionic alanine crystal have different orientations relative to the cationic ammonium moieties and thus different hydrogen bonding interactions, resulting in distinct magnetic environments, they have, consequently, distinct NMR parameters.

Overall, we observed good agreement between the experimental and calculated values for the two enantiomers of alanine. However, some discrepancies remain between the experimental and calculated NMR parameters, even within the 95\% confidence interval. In particular, the comparison reveals disagreement in $C_Q$ and $\delta_{\text{CS}}$ for at least one site. For $C_Q$, the largest deviation occurs at site O1, where the calculated value is approximately 24\% higher for the \textit{L-}enantiomer and about 14\% higher for the \textit{D-}enantiomer. A similar trend has been reported in previous \isotope[17]{O} studies of amino acid samples.~\cite{yates_theoretical_2004}  In the case of $\delta_{\text{CS}}$,  the asymmetry parameter of the CSA tensor, the largest discrepancy is again observed at site O1, with calculated values 63\% higher for  \textit{L-}alanine and 50\% higher for \textit{D-}alanine. Since NMR parameters are sensitive to the chemical environments, the weaker H-bonding environment at site O1---involved in only one \ce{C=O\cdots H-N} hydrogen bond~\cite{wu_two-dimensional_2001}---likely contributes to the observed deviation in the DFT-calculated values.~\cite{kongsted2008accuracy}

Table~\ref{tab:alaine results} also presents values of $C_Q$, $\eta_{Q}$, and $\delta_{\text{iso}}$ from previous experimental studies using powder samples. The SCNMR experimental $\eta_Q$ values obtained from this study show good agreement with the MAS~\cite{pike_solid-state_2004} and DAS~\cite{yamauchi_17o_1999} measurements. However, there is notable disagreement with the MQMAS measurements~\cite{wu_two-dimensional_2001} for \textit{D-}alanine, though the MQMAS measurements also differ from other studies as well as from the computational findings in this study. In contrast, the DFT calculated NMR parameters reported here are in good agreement with most of the experimental results. 

 The $C_Q$ and $\delta_{\text{iso}}$ have been identified as reliable parameters to assign sites O1 and O2~\cite{wu_two-dimensional_2001}. Given the nearly identical and relatively large uncertainties in experimental $C_Q$ values, and our inability to determine 95\% confidence interval for $\delta_{\text{iso}}$ in one of the two \textit{L-}alanine sites,  we have turned to the asymmetry parameter, $\eta_Q$, for the site assignment. This parameter is particularly useful, as it is known to increase with the strength of hydrogen bonding~\cite{yates_theoretical_2004}.

The Euler angles $a, b, \text{and}\ c$ agree well for both enantiomers when compared with the calculated values. Specifically, the angles $b\ \text{and}\ c$ align well with the computed value for both enantiomers. The other angle, $a$ shows the largest statistical uncertainty, which is related to the inaccurate measurement of $\eta_{\text{CS}}$~\cite{vosegaard_quadrupole_1996}, with experimental $\eta_{\text{CS}}$ values having variation between $11\%\ \text{and}\ 100\%$ in statistical uncertainty. Due to the potential ambiguity in the projection direction of the quadrupolar and CSA principal axes, the Euler angles have been rotated by $180^\circ$ where necessary to allow direct comparison. Additionally, the set of Euler angles that relate the quadrupolar PAF to both the goniometer and crystal frames has been calculated and is provided in the Supplementary Material.

\begin{figure}
    \centering
    \includegraphics[width=\linewidth]{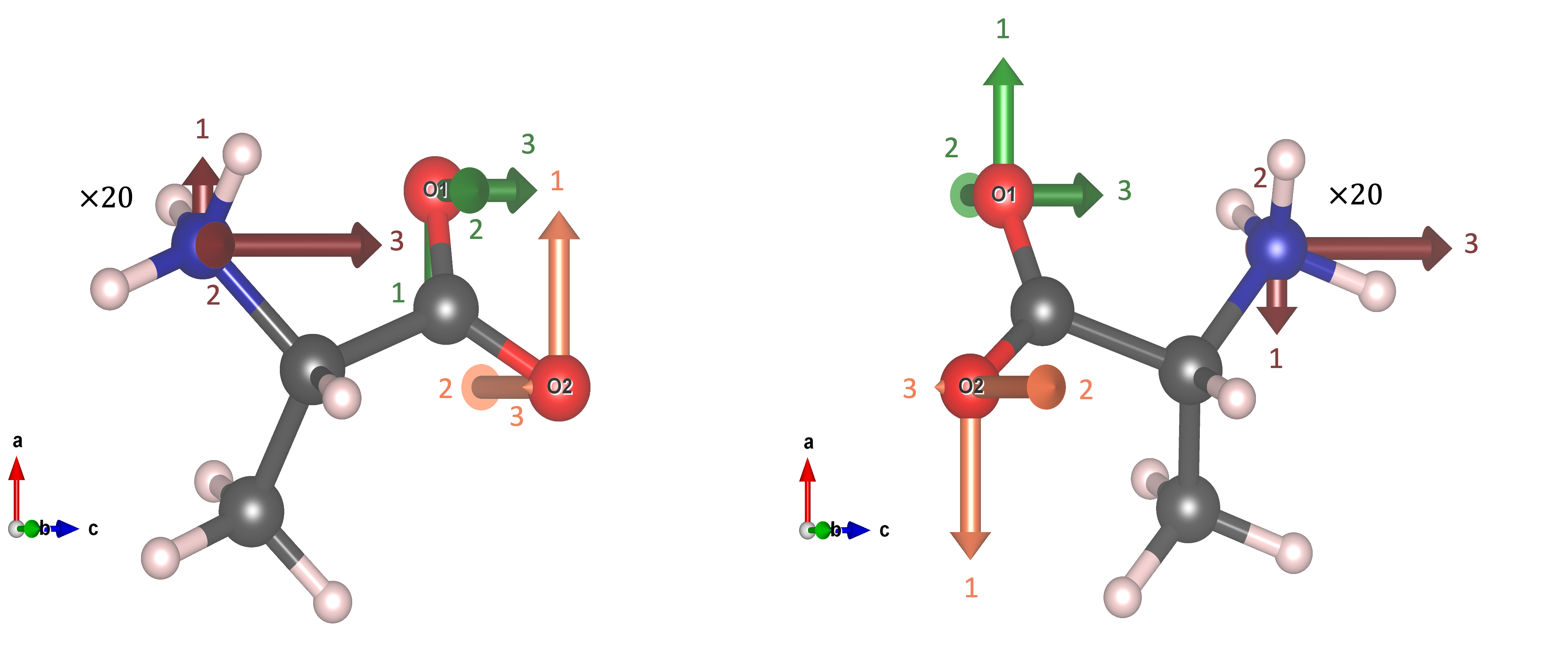}
    \caption{Magnitudes and directions of the pseudovectors for the oxygen and nitrogen sites in \textbf{Left:} \textit{L-}alanine and \textbf{Right:} \textit{D-}alanine, where $V_x = 1$, $V_y = 2$, and $V_z = 3$.  The tensors at the N site are scaled by a factor of 20 for visual clarity. In \textit{D-}alanine, the $V_y$ pseudovector of N  is hidden by the atom.  C, N, O, and H atoms are shown in gray, blue, red, and white, respectively.
    }
    \label{fig:Alanine CSA Tensor}
\end{figure}

In addition to the conventional CSA and quadrupolar tensor values, our DFT calculations also yielded the antisymmetric components of the chemical shift tensors for both \isotope[17]{O} and \isotope[14]{N} nuclei, as shown in Tables 3–5. These tables reveal that the antisymmetric chemical shift tensor components of \isotope[17]{O} and \isotope[14]{N} nuclei exhibit relationships such as $\sigma_{xz}^{\text{ACS}}(L) = -\sigma_{xz}^{\text{ACS}}(R)\  \text{and\ }   \sigma_{yz}^{\text{ACS}}(L) = - \sigma_{yz}^{\text{ACS}}(R)$ between the two enantiomers.  This confirms the mirror reflection symmetry relating the electronic environments of the two enantiomers across the xy plane. To depict the computed ACS tensors for the alanine enantiomers in their respective crystal frames, we employ the pseudovector components resulting from the contraction of ACS tensor $\bm{\sigma}^{\text{ACS}}$ and Levi-Civita symbol $\varepsilon_{ijk}$ according to Eqs. \ref{eq:Levi-Civita} and \ref{eq:pseudo_components}:~\cite{harris_tensor_2014}
\begin{equation}
    \label{eq:Levi-Civita}
    V_i = \varepsilon_{ijk}\sigma^{\text{ACS}}_{jk}
\end{equation}
giving,
\begin{align}
    \label{eq:pseudo_components}
    V_x = 2\sigma^{\text{ACS}}_{yz}, \quad
    V_y = 2\sigma^{\text{ACS}}_{zx}, \quad
    V_z = 2\sigma^{\text{ACS}}_{xy}
\end{align} Figure~\ref{fig:Alanine CSA Tensor} shows the pseudovectors that are dual of the DFT computed ACS tensors  for site O1, O2 and N, highlighting the mirror plane symmetries of the \textit{L-} and \textit{D-}enantiomers, and demonstrating the antisymmetry of the ACS tensors in chiral systems.

\begingroup
\renewcommand{\cite}[1]{\citenum{#1}}  
\begin{landscape}
\begin{table}[H]
    \centering
    \begin{adjustbox}{width=\columnwidth,center}
    \begin{tabular}{c c c c c c c c c c c c c}
            \toprule
    Compound & Nucleus & Site & $\abs{C_Q}$ / MHz & $\eta_Q$ & $\delta_{\text{CS}}$ / ppm & $\eta_{\text{CS}}$ & $\delta_{\text{iso}}\textsuperscript{\textdagger}$ / ppm & $a$ /deg  & $b$ /deg & $c$ /deg & Reference & Method\Tstrut\
    \Bstrut\\
    \midrule
    \multicolumn{1}{c}{
              \textbf{Experimental}
              }\Tstrut\Bstrut\\
 \textit{L-}alanine & \isotope[17]{O} & 1 & $6.5 \pm\, 0.4$ & $0.34 \pm\,  0.16$ & $173 \pm\, 27$ & $0.4 \pm\,0.4$ & $227^a$ & $2.3^a$ & $86 \pm\, 3$ & $92 \pm\, 14$ & \multirow{2}{10mm}{this study} &  \small SCNMR\\
\vspace{1 mm}
 
& &  2 & $6.6 \pm\, 0.4$ & $0.63 \pm\,  0.17$ & $189 \pm\, 22$ & $0.5 \pm\,0.2$ & $238.8 \pm\,7.8$ & $32 \pm\, 16$ & $86 \pm\, 3$ & $89 \pm\, 5$ \\
\\
 & \isotope[17]{O} & 1 & $7.86 \pm\, 0.05$ & $0.28 \pm\,  0.02$ & - & - & $284.0 \pm\,0.5$ & - & - & - & \multirow{2}{10mm}{~\cite{pike_solid-state_2004}} &  \small MAS\\
\vspace{1 mm}
& & 2 & $6.53 \pm\, 0.05$ & $0.70 \pm\,  0.02$ & - & - & $260.5 \pm\,0.5$  & - & - & -\\
\\
  & \isotope[17]{O} & 1 & $8.1 \pm\, 0.3$ & - & - & - & $285 \pm\,8$ & - & - & - & \multirow{2}{10mm}{~\cite{yamauchi_17o_1999, pike_solid-state_2004}} &  \small DAS\\
\vspace{1 mm}
& & 2 & $7.2 \pm\, 0.3$ & - & - & - & $268 \pm\,8$  & - & - & -\\
\\
\textit{D-}alanine & \isotope[17]{O} & 1 & $7.1	\pm\, 0.2$ & $0.24 \pm\,0.09$ & $188	\pm\,15$ & $0.4	\pm\,0.12$ & $235.5 \pm\,4.9$ & $24 \pm\,10$ & $89.1 \pm\,1.5$ & $86 \pm\,10$ & \multirow{2}{10mm}{this study} &  \small SCNMR\\
\vspace{1 mm}
& & 2 & $5.91 \pm\, 0.13$ & $0.70 \pm\,  0.06$ & 1$53 \pm\, 7$ & $0.83 \pm\,0.09$ & $224.8 \pm\,2.8$ & $27 \pm\, 3$ & $88.1 \pm\, 1.0$ & $91.0 \pm\, 1.3$\\
\\
 & \isotope[17]{O} & 1 & $7.60 \pm\, 0.02$ & $0.60 \pm\,  0.01$ & - & - & $275 \pm\,5$ & - & - & - & \multirow{2}{10mm}{~\cite{wu_two-dimensional_2001}} &  \small MQMAS\\
\vspace{1 mm}
& & 2 & $6.40 \pm\, 0.02$ & $0.65 \pm\,  0.01$ & - & - & $262 \pm\,5$  & - & - & -\\

 \multicolumn{1}{c}{\textbf{Calculated}} \\
                
                 \multicolumn{1}{c}{\footnotesize \textbf{PBEsol + GIPAW}}\\
            
                \textit{L-} and \textit{D-}alanine & \isotope[17]{O} & 1 & 8.09 & 0.22 & 282 & 0.48 & $-30.7^b$ & 39 & 89 & 97 & \multirow{2}{10mm}{this study} & -\\
                \vspace{1 mm}
                & & 2 & 6.49 & 0.65 & 202 & 0.68 & $-11.9^b$ & 29 & 88 & 95\\
               \\
               \cmidrule(lr){2-11}
                & \isotope[14]{N} & - & 1.29 & 0.25 & 10.4 & 0.99 & $186.4^b$ & 125 & 150 & 150 & \multirow{2}{10mm}{this study} & -\\
                \\
     \bottomrule
    \end{tabular} 
    \end{adjustbox}
    \begin{minipage}{\linewidth}
\footnotesize
\textsuperscript{\textdagger} Isotropic chemical shifts are relative to tap water at 0.0 ppm.\\
$^a$ Error limit could not be estimated for the parameter.\\
$^b$ Unreferenced
\end{minipage}
    \caption{Experimental and computed quadrupolar couplings, chemical shift anisotropies, isotropic chemical shift, and relative orientations of the two tensors for \isotope[17]{O} and \isotope[14]{N} nuclei in  \textit{L-}alanine and \textit{D-}alanine.}
    \label{tab:alaine results}

\end{table}
\end{landscape}
\endgroup


\section{Conclusion}

In this study, we investigated the quadrupolar and chemical shielding tensor components of \isotope[17]{O}-enriched alanine enantiomers using single-crystal ssNMR spectroscopy. Eight magnetically inequivalent \isotope[17]{O} sites, as predicted by X-ray crystallography, were successfully identified and analyzed for their NMR tensor parameters. The experimental analysis performed using the ASICS software package, as described in the~\nameref{section: alanine analysis} section, was further supported by DFT calculations. The DFT calculations not only reproduced the experimental tensor parameters with good accuracy but also aided in assigning the crystallographically nonequivalent \isotope[17]{O} sites, thereby validating the spectroscopic findings.

This study provides, for the first time, an extended and detailed characterization of the crystallographically distinct \isotope[17]{O} nuclei in alanine enantiomers. The obtained NMR parameters are in good agreement with previous studies, reinforcing the robustness of our approach. Although the antisymmetric chemical shift (ACS) tensor components could not be directly extracted from the ssNMR experiments—since only the central transition of the spin-5/2 \isotope[17]{O} nucleus was analyzed, which lacks ACS contributions—the DFT results revealed the presence of off-diagonal ACS elements (e.g., $\sigma_{yz}^{\text{ACS}} = -\sigma^{\text{ACS}}_{zy}$) in the crystal frame. These components reflect the mirror-symmetric electronic environments characteristic of the optical isomers.

Our findings demonstrate the utility of combining single-crystal NMR spectroscopy with DFT calculations to discern subtle differences in electronic environments between enantiomers. This capability underscores the power of single-crystal NMR for probing chiral molecular systems such as amino acids. Future studies may focus on analyzing all satellite transitions of \isotope[17]{O} nuclei in chiral environments to experimentally access ACS contributions through Quadrupolar–ACS interactions~\cite{wi_quadrupolar-shielding_2002}.  Such interactions can produce observable features in the angular-dependent rotation patterns of single-crystal ssNMR spectra, in addition to the conventional first- and second-order quadrupolar and CSA effects. Furthermore, we propose extending this methodology to \isotope[14]{N} single-crystal NMR at higher magnetic fields, where enhanced sensitivity and resolution may enable clearer observation of quadrupolar broadening and magnetic shielding anisotropy. These studies could provide deeper insights into antisymmetric shielding effects and their potential role in chiral selection mechanisms, offering a plausible explanation for the enantiomeric excess of \textit{L-}amino acids observed in carbonaceous chondrite meteorites—and thus contributing to our understanding of prebiotic chemistry in astrophysical environments.

\begin{acknowledgement}

A portion of this work was performed at the National High Magnetic Field Laboratory (NHMFL), which is supported by National Science Foundation Cooperative Agreement No. DMR-2128556 and the State of Florida.
We  thank the Gordon and Betty Moore Foundation for financially supporting this research through Grant GBMF7799 to Western Michigan University as well as the NASA Early Career Collaboration Award to S.A. that partially supported travel to NHMFL. This work was facilitated by software tools (specifically MagresPython and Soprano) developed by the Collaborative Computing Project for NMR Crystallography, funded by EPSRC grant EP/T026642/1.

\end{acknowledgement}

\section{Competing Interests}
The authors declare no competing interests.

\begin{suppinfo}

NMR spectra, x-ray orientation of \textit{D-} and \textit{L-}alanine, computational model comparison, mass spectrometric analysis spectra, and the NMR pulse program used for data collection.

\end{suppinfo}

\bibliography{antisymm_mag_enantiomers}

\end{document}